\RequirePackage{fix-cm}
\documentclass{iopjournal}
\usepackage[english]{babel}
\usepackage{amsmath}
\usepackage{amssymb}
\usepackage{amsfonts}
\usepackage{ragged2e}
\usepackage{multirow}
\usepackage{cite}
\usepackage{fourier}
\justifying

\begin{document}

\title{Symmetries and Propagation of Surface Plasmon Polaritons in Axion Electrodynamics}

\author{Marco Peluso~\orcid{0009-0009-3314-9040}}

\affil{Dipartimento di Scienza Applicata e Tecnologia, Politecnico di Torino, 10129 Torino, Italy}

\email{marco.peluso@polito.it}


\begin{abstract}
    \justifying{Surface plasmon polaritons at the interface between a conductor governed by axion electrodynamics and an ordinary dielectric are studied using a general phenomenological model, in which the macroscopic response is described by an axial vector $ \mathbf{G} $, responsible for the anomalous Hall effect, and a pseudoscalar $ W $, accounting for the chiral magnetic effect. We extend the analysis of the chiral magnetic effect in the planar geometry, address its role in the cylindrical one and, in both geometries, discuss the case in which the two effects coexist. We find an upper bound on $ W $, a nonreciprocity requiring both effects and the absence of freely propagating modes, depending on the orientation of $ \mathbf{G} $, for the planar interface, and, for the cylindrical one, a novel coupling between the linear and orbital angular momenta of the modes, which already arises from the chiral magnetic effect alone and persists when the anomalous Hall effect is also present. The exact eigenmodes of the axion medium are derived, and the symmetries and propagation features of the surface modes are made explicit by analytical and asymptotic evaluation of the dispersion relations.}
\end{abstract}

\section{Introduction}
\label{sec:Introduction}

Among the various manifestations of light--matter interaction, surface plasmon polaritons (SPPs) are hybrid excitations coupling the electromagnetic field to charge oscillations, localised at the interface between a conductor and a dielectric~\cite{Barnes2003,Maier2007}. Being confined to regions much smaller than the wavelength, SPPs strongly enhance the local field intensity~\cite{Yu2019}. Moreover, SPP waveguides exploit this confinement to carry light in structures only a few tens of nanometres across, well below the diffraction limit of conventional dielectric optics~\cite{Gramotnev2010}, and are the building blocks of devices such as plasmonic nanolasers~\cite{Oulton2009,Hill2009}. Other applications require nonreciprocal propagation, in which counter-propagating modes obey different dispersion relations~\cite{Camley1987,Caloz2018}. In conventional conductors, reciprocity is usually broken by an external magnetic induction, which turns SPPs into surface magnetoplasmons~\cite{ChiuQuinn1972,Wallis1974}; sizeable effects, however, require strong fields~\cite{Bliokh2018}.

An alternative is provided by conductors governed by effective axion electrodynamics, in which reciprocity can be broken by the response of the medium itself, without any external field~\cite{Kotov2018,Guo2023}. The axion field was originally introduced in particle physics to address the strong CP problem~\cite{PecceiQuinn1977,Weinberg1978,Wilczek1978}. Its coupling to the electromagnetic field modifies Maxwell's theory through a term proportional to $ \mathbf{E} \cdot \mathbf{B} $~\cite{Sikivie1983,Wilczek1987}. The same term emerges in topological insulators~\cite{Qi2008,Sekine2021} and semimetals~\cite{Zyuzin2012,Chen2013}, and in chiral plasmas~\cite{Fukushima2008,Kharzeev2014}. When the axion field varies linearly in space and time, its contribution to the response of a conductor is fixed by two quantities: a vector, from its spatial gradient, and a scalar, from its time derivative, which we denote here by $ \mathbf{G} $ and $ W $, respectively.

The vector $ \mathbf{G} $ is responsible for the anomalous Hall effect (AHE): it enters the permittivity as an antisymmetric, gyrotropic term and acts as an effective intrinsic magnetic field~\cite{Burkov2014,Hofmann2016}. The scalar $ W $ instead generates a current parallel to the magnetic induction, the chiral magnetic effect (CME)~\cite{Fukushima2008,Kharzeev2014}, and enters the wave equation through the curl of the electric field, with the same structure as natural optical activity in chiral media~\cite{Landau1984,Ma2015}. Such a conductor therefore combines, with independent strengths, both effects. In crystals, a static chiral magnetic current cannot flow in equilibrium~\cite{Vazifeh2013}, but the dynamical response at finite frequency survives~\cite{Chen2013,Ma2015,Zhong2016} and a chiral imbalance can be sustained out of equilibrium~\cite{Son2013,Li2016}. It is also through the CME that axion dark matter would manifest itself in a conductor~\cite{Hong2022,Bose2026}. A homogeneous, time-dependent axion background converts into photons in the presence of a strong static magnetic induction, which is the operating principle of haloscopes~\cite{Sikivie1983,Irastorza2018}. In Weyl semimetals, the axion--electron coupling instead induces a chiral chemical potential proportional to the time derivative of the axion field, and the resulting chiral magnetic current has been proposed as a direct probe of dark matter~\cite{Irastorza2018,Bose2026}. Detection schemes based on condensed-matter axion quasiparticles follow a similar idea~\cite{Marsh2019,SchutteEngel2021}.

Within axion electrodynamics, the study of SPPs has so far focused almost exclusively on the AHE. Following the terminology of magneto-optics~\cite{ChiuQuinn1972,Wallis1974,Camley1987}, where the Faraday and Voigt geometries refer to an applied magnetic induction parallel and orthogonal to the direction of propagation, three inequivalent orientations of the axion gradient are usually considered at a planar interface: Voigt, in which it lies in the plane of the interface and is orthogonal to the wavevector; Faraday, where it still lies in the plane of the interface but is parallel to the wavevector; and perpendicular, in which it is normal to the interface. SPPs are found to be nonreciprocal in the Voigt case and reciprocal in the other two~\cite{Hofmann2016}; this analysis was extended to thin films~\cite{Kotov2018,Tamaya2019}, strained slabs~\cite{Bugaiko2020} and slot waveguides~\cite{OskouiAbdol2019}. In cylindrical waveguides whose core is the conductor, the AHE lifts the degeneracy between modes with opposite orbital angular momentum (OAM)~\cite{Peluso2025WG}, which in conical tips results in the selective nanofocusing of all the modes winding in the same direction~\cite{Peluso2025NF}. The CME alone, on the other hand, has been considered only numerically, in the planar geometry~\cite{Hofmann2016,OskouiAbdol2020}. Together, the two effects allow bulk transverse electromagnetic modes in a hollow metallic waveguide filled with an axion medium, where the metal only provides the boundary~\cite{GomesMouraMelo2025}. By contrast, the consequences of their coexistence for surface modes remain unexplored in both the planar and the cylindrical geometry, as does the role of the CME alone in the latter.

In this paper, we address this gap by studying SPPs at the interface between a dielectric and a conductor governed by axion electrodynamics, first extending the analysis of the CME alone and then considering its coexistence with the AHE. The ordinary response of the conduction electrons is modelled by a dissipationless Drude permittivity~\cite{AshcroftMermin,Grosso2000}. We consider $ \mathbf{G} $ in each of the three configurations at the planar interface and along the axis of the cylindrical one. For both geometries, we derive the exact eigenmodes of the conductor in closed form and, from them, the exact secular equations defining the surface modes. Their asymptotic expansions, which complement the numerical solutions, make the symmetries of the dispersion relations explicit. We find an upper bound on $ W $ for the existence of SPPs. In the planar geometry, $ W $ alone can reverse the group velocity above a threshold; combined with $ \mathbf{G} $, it breaks reciprocity in the Faraday configuration, modifies the nonreciprocity of the Voigt one, and suppresses freely propagating surface modes in the perpendicular one. In the cylindrical geometry, $ W $ alone couples the linear and the orbital angular momenta of the modes, and together with $ \mathbf{G} $ it lifts all the degeneracies between the directions of propagation and rotation.

Since our aim is to establish how the AHE and the CME modify the symmetries and the propagation features of SPPs, rather than to model a specific system, we regard $ \mathbf{G} $ and $ W $ as independent phenomenological parameters, whose microscopic origin and physical realisation lie beyond the scope of this work.

The rest of the paper is organised as follows. Section~\ref{sec:Model} introduces the model and the general structure of the surface modes. The results for the planar and cylindrical interfaces are presented in Sections~\ref{sec:Planar} and~\ref{sec:Cylindrical}, respectively, while Section~\ref{sec:Conclusions} discusses these findings and ends with concluding remarks. Technical details of the calculations are collected in the Appendices.

\begin{figure}[htbp]
    \centering
    \includegraphics[width=0.9 \textwidth]{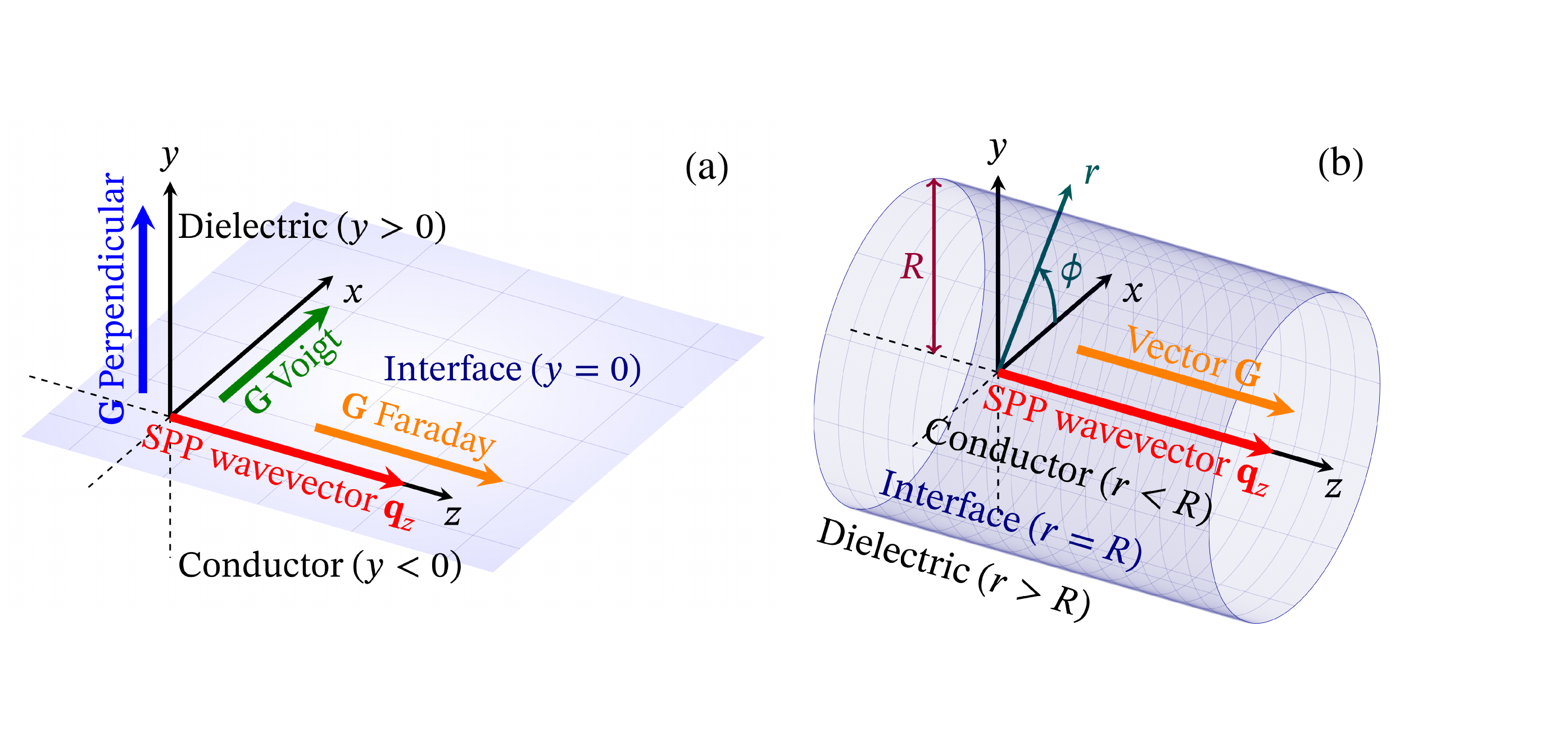}
\caption{%
    \justifying Geometries and configurations.
    Panel \textbf{(a)}: planar interface $ y = 0 $ between the conductor ($ y < 0 $) and the dielectric ($ y > 0 $). The SPP propagates along $ z $ with wavevector $ q_z \mathbf{e}_z $, and $ \mathbf{G} $ lies along $ x $ (Voigt), $ z $ (Faraday) or $ y $ (perpendicular configuration).
    Panel \textbf{(b)}: cylindrical interface $ r = R $ between the conductor ($ r < R $) and the dielectric ($ r > R $), with $ q_z \mathbf{e}_z $ and $ \mathbf{G} $ along the axis $ z $; $ r $ and $ \phi $ are the radial and azimuthal coordinates.
}
    \label{fig:figure1}
\end{figure}

\section{Model}
\label{sec:Model}

We describe the bulk of the conductor through an effective action in which the electromagnetic field, coupled to the electronic charge and current, is supplemented by an axion term proportional to $ \mathbf{E} \cdot \mathbf{B} $~\cite{Wilczek1987,Carroll1990,Sekine2021}. In SI units, it reads
\begin{equation}\label{eq:action}
    S_{\mathrm{AE}} = \frac{ 1 }{ 2 \mu_0 } \int \left( \frac{ \mathbf{E} ^2 }{ c ^2 } - \mathbf{B} ^2 \right) \, \mathrm{d} ^3 x \, \mathrm{d} t - \int \left( \rho \, \Phi - \mathbf{J} \cdot \mathbf{A} \right) \, \mathrm{d} ^3 x \, \mathrm{d} t + \varepsilon_0 \, c \int \left( W c t - \mathbf{G} \cdot \mathbf{x} \right) \, \left( \mathbf{E} \cdot \mathbf{B} \right) \, \mathrm{d} ^3 x \, \mathrm{d} t ,
\end{equation}
where $ \rho $ and $ \mathbf{J} $ are the ordinary densities of electronic charge and current, the scalar potential $ \Phi $ and the vector potential $ \mathbf{A} $ are such that $ \mathbf{E} = - \nabla \Phi - \partial_t \mathbf{A} $ and $ \mathbf{B} = \nabla \times \mathbf{A} $, and $ c ^2 = 1 / \left( \varepsilon_0 \mu_0 \right) $.

The quantities $ \mathbf{G} $ and $ W $, both with the dimensions of an inverse length, are the gradient and the time derivative of the axion field, respectively, and are taken to be real and constant throughout the conductor. Their transformation properties follow from the invariance of the action. Under time reversal, $ \mathbf{E} $ is even and $ \mathbf{B} $ is odd, and therefore $ \mathbf{E} \cdot \mathbf{B} $ changes sign, and the action is invariant only if the axion field $ W c t - \mathbf{G} \cdot \mathbf{x} $ changes sign too; since $ t \to - t $, this requires $ \mathbf{G} \to - \mathbf{G} $ and $ W \to W $. Under parity, $ \mathbf{E} $ is odd and $ \mathbf{B} $ is even, and the same argument, with $ \mathbf{x} \to - \mathbf{x} $, gives $ \mathbf{G} \to \mathbf{G} $ and $ W \to - W $. Hence $ \mathbf{G} $ is an axial vector, odd under time reversal, and $ W $ is a pseudoscalar, even under time reversal and odd under parity.

Varying Eq.~\eqref{eq:action} with respect to $ \Phi $ and $ \mathbf{A} $ gives the inhomogeneous Maxwell equations (Gauss's law and Amp\`ere--Maxwell law), while the homogeneous ones (Gauss's law for magnetism and Faraday's law) follow identically from the definition of the potentials. The macroscopic electromagnetic response of the conductor is therefore governed by
\begin{align}
    \nabla \cdot \mathbf{E} &= \frac{ \rho }{ \varepsilon_{0} } + c \, \mathbf{G} \cdot \mathbf{B}, \label{eq:GaussE_total} \\[6pt]
    \nabla \cdot \mathbf{B} &= 0, \label{eq:GaussB_total} \\[6pt]
    \nabla \times \mathbf{E} &= - \frac{ \partial \mathbf{B} }{ \partial t }, \label{eq:Faraday_total} \\[6pt]
    \nabla \times \mathbf{B} - \frac{ 1 }{ c ^2 } \frac{ \partial \mathbf{E} }{ \partial t } &= \mu_{0} \mathbf{J} - \frac{ 1 }{ c } \mathbf{G} \times \mathbf{E} + W \mathbf{B}. \label{eq:Amp-Max_total}
\end{align}
The axion term adds the anomalous charge density $ \varepsilon_0 c \, \mathbf{G} \cdot \mathbf{B} $ to Gauss's law and the anomalous current density $ - \varepsilon_0 c \, \mathbf{G} \times \mathbf{E} + W \mathbf{B} / \mu_0 $ to the Amp\`ere--Maxwell law. The first contribution to the current is the anomalous Hall current, orthogonal to both $ \mathbf{G} $ and $ \mathbf{E} $; the second is the chiral magnetic current, parallel to $ \mathbf{B} $~\cite{Zyuzin2012,Chen2013}. The anomalous charge and current densities satisfy the continuity equation identically, as a consequence of Faraday's law. We consider monochromatic fields with time dependence $ \mathit{e} ^{ - \mathrm{i} \omega t } $ and a local and isotropic response, for which the current density and, through the continuity equation $ \nabla \cdot \mathbf{J} = \mathrm{i} \omega \rho $, the charge density read
\begin{align}\label{eq:currents}
    \mathbf{J} &= - \mathrm{i} \omega \varepsilon_0 \left[ \epsilon \left( \omega \right) - 1 \right] \mathbf{E}, &
    \rho &= - \varepsilon_0 \left[ \epsilon \left( \omega \right) - 1 \right] \nabla \cdot \mathbf{E} .
\end{align}

For the relative permittivity we take the Drude form
\begin{equation}\label{eq:Drude}
    \epsilon \left( \omega \right) = \epsilon_\infty \left[ 1 - \left( \frac{ \omega_p }{ \omega } \right) ^2 \right] ,
\end{equation}
where $ \epsilon_\infty $ is the background dielectric constant and $ \omega_p $ is the plasma frequency, defined as the zero of $ \epsilon \left( \omega \right) $~\cite{AshcroftMermin,Grosso2000}. Eliminating $ \mathbf{B} $ through Faraday's law, and $ \rho $ and $ \mathbf{J} $ through Eq.~\eqref{eq:currents}, the Amp\`ere--Maxwell law yields the wave equation for the electric field in the conductor,
\begin{equation}
    \nabla \left( \nabla \cdot \mathbf{E} \right) - \nabla ^2 \mathbf{E} - \frac{ \omega ^2 }{ c ^2 } \epsilon \left( \omega \right) \, \mathbf{E} + \frac{ \mathrm{i} \omega }{ c } \mathbf{G} \times \mathbf{E} - W \nabla \times \mathbf{E} = \mathbf{0} . \label{eom}
\end{equation}

The term in $ \mathbf{G} $ is an antisymmetric contribution to the permittivity tensor, with the same gyrotropic structure as in a magnetised plasma, $ \mathbf{G} $ playing the role of the gyration vector~\cite{Landau1984}. The term in $ W $ involves the curl of the field and is therefore a first-order spatial dispersion~\cite{Landau1984,Ma2015}. Both are collected in the displacement field of the conductor, given in Appendix~\ref{app:Boundary}. Combining Gauss's law~\eqref{eq:GaussE_total} with Eq.~\eqref{eq:currents} and with Faraday's law gives the transversality condition
\begin{equation}
    \nabla \cdot \left[ \epsilon \left( \omega \right) \mathbf{E} + \frac{ c }{ \mathrm{i} \omega } \mathbf{G} \times \mathbf{E} \right] = 0 , \label{divE}
\end{equation}
which replaces $ \epsilon \left( \omega \right) \nabla \cdot \mathbf{E} = 0 $ of a conventional conductor. The scalar $ W $ does not appear in it, because the chiral magnetic current is divergence-free. For a plane wave $ \mathbf{E} \propto \mathit{e} ^{ \mathrm{i} \mathbf{q} \cdot \mathbf{x} } $, Eq.~\eqref{eom} has nontrivial solutions only if
\begin{equation}\label{eq:bulk_imp_dispersion_relation}
    \left[ \left( \mathbf{q} \cdot \mathbf{q} \right) - \frac{ \omega ^2 }{ c ^2 } \epsilon \left( \omega \right) \right] ^2 \epsilon \left( \omega \right) + \| \mathbf{q} \times \mathbf{G} \| ^2 - \left( \mathbf{G} \cdot \mathbf{G} \right) \frac{ \omega ^2 }{ c ^2 } \epsilon \left( \omega \right) - \left[ W ^2 \left( \mathbf{q} \cdot \mathbf{q} \right) - 2 \frac{ \omega }{ c } W \left( \mathbf{q} \cdot \mathbf{G} \right) \right] \epsilon \left( \omega \right) = 0 ,
\end{equation}
defining the bulk dispersion relation of the conductor. For $ \mathbf{G} = \mathbf{0} $ and $ W = 0 $, it reduces to that of a conventional conductor $ \mathbf{q} \cdot \mathbf{q} = \omega ^2 \epsilon \left( \omega \right) / c ^2 $. The vector $ \mathbf{G} $ enters through $ \| \mathbf{q} \times \mathbf{G} \| ^2 $ and $ \mathbf{G} \cdot \mathbf{G} $, which are even in both $ \mathbf{q} $ and $ \mathbf{G} $ and make the response anisotropic, while $ W $ alone enters through $ W ^2 \left( \mathbf{q} \cdot \mathbf{q} \right) $, which is isotropic and even in $ W $. The term $ W \left( \mathbf{q} \cdot \mathbf{G} \right) $ is the only one odd in $ \mathbf{q} $ and requires both effects: it reflects the fact that $ W \mathbf{G} $ transforms as $ \mathbf{q} $ under parity and time reversal.

The dielectric is taken to be an ordinary, nonmagnetic and isotropic medium. Maxwell's equations in the dielectric follow from the ones above by setting $ \mathbf{G} = \mathbf{0} $ and $ W = 0 $, and replacing $ \epsilon \left( \omega \right) $ with $ \epsilon_d $, a positive, frequency-independent constant.

At the interface between the two media, we consider SPPs propagating along the $ z $ axis with real wavevector $ q_z $, in the two geometries sketched in Fig.~\ref{fig:figure1}. The planar interface is the plane $ y = 0 $, with the conductor at $ y < 0 $ and the dielectric at $ y > 0 $. We take $ \mathbf{G} $ along one of the Cartesian unit vectors $ \mathbf{e}_x $, $ \mathbf{e}_z $ or $ \mathbf{e}_y $, corresponding to the Voigt, Faraday and perpendicular configurations, respectively. The cylindrical interface is the surface of the conductor, a cylinder of radius $ R $, surrounded by the dielectric, with $ \mathbf{G} $ along its axis.

In the planar picture, we take the fields to be uniform along $ x $, and in the dielectric they are then a superposition of a transverse electric (TE) and a transverse magnetic (TM) evanescent wave, with $ \mathbf{q} = \left( 0 , \mathrm{i} \kappa_d , q_z \right) $ and inverse decay length $ \kappa_d $ fixed by the dispersion relation $ \mathbf{q} \cdot \mathbf{q} = \omega ^2 \epsilon_d / c ^2 $. For the field to be localised at the interface, $ \kappa_d $ must have a positive real part, which holds outside the light cone of the dielectric, defined by
\begin{equation}\label{eq:lightcone}
    \omega \geq \frac{ c \, | q_z | }{ \sqrt{ \epsilon_d } } ,
\end{equation}
on whose boundary, the dielectric light line, $ \kappa_d $ vanishes. In the conductor, the anomalous terms couple the two polarisations, and the field is a superposition of the two eigenmodes of Eq.~\eqref{eom}, whose inverse decay lengths $ \kappa_\pm $, complex in general and with $ \operatorname{Re} \left( \kappa_\pm \right) > 0 $, are the roots of Eq.~\eqref{eq:bulk_imp_dispersion_relation} for $ \mathbf{q} = \left( 0 , - \mathrm{i} \kappa_\pm , q_z \right) $. The fields in the two media thus read
\begin{align}
    \mathbf{E} \left( y > 0 \right) &= \left( \widetilde{\mathbf{E}}_{\mathrm{TEd}} + \widetilde{\mathbf{E}}_{\mathrm{TMd}} \right) \mathit{e} ^{ - \kappa_d y + \mathrm{i} q_z z } , \\[8pt]
    \mathbf{E} \left( y < 0 \right) &= \left( \widetilde{\mathbf{E}}_+ \mathit{e} ^{ \kappa_+ y } + \widetilde{\mathbf{E}}_- \mathit{e} ^{ \kappa_- y } \right) \mathit{e} ^{ \mathrm{i} q_z z } ,
\end{align}
where the exact closed-form expressions of the amplitude vectors, which depend on the configuration, are given in Appendices~\ref{app:PlanarDiel} and~\ref{app:PlanarCond}.

In the cylindrical picture, $ \mathbf{G} \parallel \mathbf{e}_z $ preserves the rotational symmetry about the cylinder axis, so that the field can be expanded in eigenstates of the OAM, labelled by $ m \in \mathbb{Z} $. The radial dependence is given by the modified Bessel functions $ K_m $ in the dielectric and $ I_m $ in the conductor, evaluated at $ \kappa_d r $ and $ \kappa_\pm r $, respectively, where we keep the notation $ \kappa_\pm $ for the radial inverse decay lengths:
\begin{align}
    \mathbf{E} \left( r > R \right)
    &= \sum_{ m = - \infty } ^{ + \infty }
    \left[ \widetilde{\mathbf{E}}_m ^{\mathrm{I}} \left( \kappa_d r \right) + \widetilde{\mathbf{E}}_m ^{\mathrm{II}} \left( \kappa_d r \right) \right]
    \mathit{e} ^{ \mathrm{i} \left( m \phi + q_z z \right) } , \\[8pt]
    \mathbf{E} \left( r < R \right)
    &= \sum_{ m = - \infty } ^{ + \infty }
    \left[ \widetilde{\mathbf{E}}_m ^+ \left( \kappa_+ r \right) + \widetilde{\mathbf{E}}_m ^- \left( \kappa_- r \right) \right]
    \mathit{e} ^{ \mathrm{i} \left( m \phi + q_z z \right) } ,
\end{align}
where the exact closed-form expressions of the amplitude vectors are given in Appendices~\ref{app:CylindricalDiel} and~\ref{app:CylindricalCond}.

The fields in the two media are then matched through the boundary conditions at the interface. We treat the anomalous terms as bulk contributions, as done in Refs.~\cite{Hofmann2016,Kotov2018}; the interface carries no surface charges or currents, and the matching conditions take the standard form (see Appendix~\ref{app:Boundary}): the tangential components of the electric field and of the magnetic field $ \mathbf{H} = \mathbf{B} / \mu_0 $ are continuous,
\begin{align}\label{eq:boundary}
    \Delta \mathbf{E}_\parallel &= \mathbf{0} , &
    \Delta \mathbf{H}_\parallel &= \mathbf{0} ,
\end{align}
where $ \Delta $ denotes the difference between the values on the two sides of the interface.

These conditions give four linear and homogeneous equations for the four amplitudes of the fields, separately for each $ m $ in the cylinder. The system admits a nontrivial solution only if the determinant of the corresponding coefficient matrix $ \mathcal{M} $ vanishes:
\begin{equation}\label{eq:secular}
    \det{ \mathcal{M} \left( \omega , q_z , \mathbf{G} , W \right) } = 0 .
\end{equation}
This is the secular equation of the surface modes, with $ m $ and $ R $ as additional parameters for the cylinder. It implicitly defines the dispersion relation $ \omega \left( q_z \right) $, which we denote by $ \omega_m \left( q_z \right) $ when it depends on the OAM. In general, the secular equation must be solved numerically. In asymptotic limits, such as $ | q_z | \to \infty $ or $ | m | \to \infty $, however, the roots can be expanded analytically as described in Appendix~\ref{app:Math}. These expansions are the main tool of Sections~\ref{sec:Planar} and~\ref{sec:Cylindrical}, as they make explicit how $ \mathbf{G} $ and $ W $ enter the dispersion relations and thus reveal the symmetries and propagation features of the modes.

The solutions of Eq.~\eqref{eq:secular} describe SPPs only where the fields decay on both sides of the interface. In addition to the dielectric light cone~\eqref{eq:lightcone}, the anomalous terms give rise to a forbidden region of the $ \left( q_z , \omega \right) $ plane, defined by
\begin{equation}\label{eq:forbiddenregion}
    \min \left\{ \operatorname{Re} \left( \kappa_+ \right) , \operatorname{Re} \left( \kappa_- \right) \right\} = 0 ,
\end{equation}
where the SPPs are no longer bound to the interface and hybridise with the bulk modes of the conductor. This region is shaded in dark grey in Figs.~\ref{fig:figure2}--\ref{fig:figure6} and Fig.~\ref{fig:figure8}. The numerical solutions shown in the figures are obtained for $ \epsilon_d = \epsilon_\infty = 1 $.

\section{Planar Interface}
\label{sec:Planar}

At the planar interface of Fig.~\ref{fig:figure1}(a), we write $ \mathbf{G} = G_{\mathrm{V}} \mathbf{e}_x $, $ G_{\mathrm{F}} \mathbf{e}_z $ or $ G_{\mathrm{P}} \mathbf{e}_y $, in the Voigt, Faraday and perpendicular configurations, respectively. We first recall the case of the AHE alone, then consider the CME alone and finally the coexistence of the two effects in the three configurations.

\subsection{AHE alone}
\label{sec:PlanarAHE}

For $ \mathbf{G} \neq \mathbf{0} $ and $ W = 0 $, the problem reduces to the one first studied in Ref.~\cite{Hofmann2016}, which we briefly review to fix the notation and to provide the corresponding analytical expansions of the dispersion relations. With respect to conventional SPPs (see Appendix~\ref{app:PlanarDiel}), the main new feature is nonreciprocity, which appears in the Voigt configuration~\cite{Hofmann2016} and is made explicit by the large-wavevector expansion $ c | q_z | \gg \omega_p $ of the dispersion relation,
\begin{equation}
    \omega_{\mathrm{V}} \left( q_z \to \pm \infty \right) = \omega_{\infty,s} ^{\mathrm{V}} - \frac{ \epsilon_d ^2 \, \left( \omega_{\infty,s} ^{\mathrm{V}} \right) ^4 }{ c ^2 \sqrt{ c ^2 G_{\mathrm{V}} ^2 + 4 \left( \epsilon_\infty + \epsilon_d \right) \epsilon_\infty \omega_p ^2 } } \frac{ 1 }{ q_z ^2 } + \omega_p \, \mathcal{O} \left( \frac{ \omega_p ^4 }{ c ^4 q_z ^4 } \right) ,
\end{equation}
where
\begin{equation}\label{eq:asymVoigt}
    \omega_{\infty,s} ^{\mathrm{V}} \left[ \operatorname{sgn} \left( q_z \right) , G_{\mathrm{V}} \right] = \frac{ \sqrt{ c ^2 G_{\mathrm{V}} ^2 + 4 \left( \epsilon_\infty + \epsilon_d \right) \epsilon_\infty \omega_p ^2 } + \operatorname{sgn} \left( q_z \right) \, c \, G_{\mathrm{V}} }{ 2 \left( \epsilon_\infty + \epsilon_d \right) }
\end{equation}
are the large-wavevector asymptotes~\cite{OskouiAbdol2020}, with $ s = \operatorname{sgn} \left( q_z \right) $.
The two are separated by $ c \, G_{\mathrm{V}} / \left( \epsilon_\infty + \epsilon_d \right) $, linearly in $ G_{\mathrm{V}} $, and when it vanishes both reduce to the surface plasmon frequency
\begin{equation}\label{eq:omegainfty}
    \omega_\infty = \omega_p \sqrt{ \frac{ \epsilon_\infty }{ \epsilon_\infty + \epsilon_d } }.
\end{equation}

\begin{figure}[htbp]
    \centering
    \includegraphics[width=0.45\textwidth]{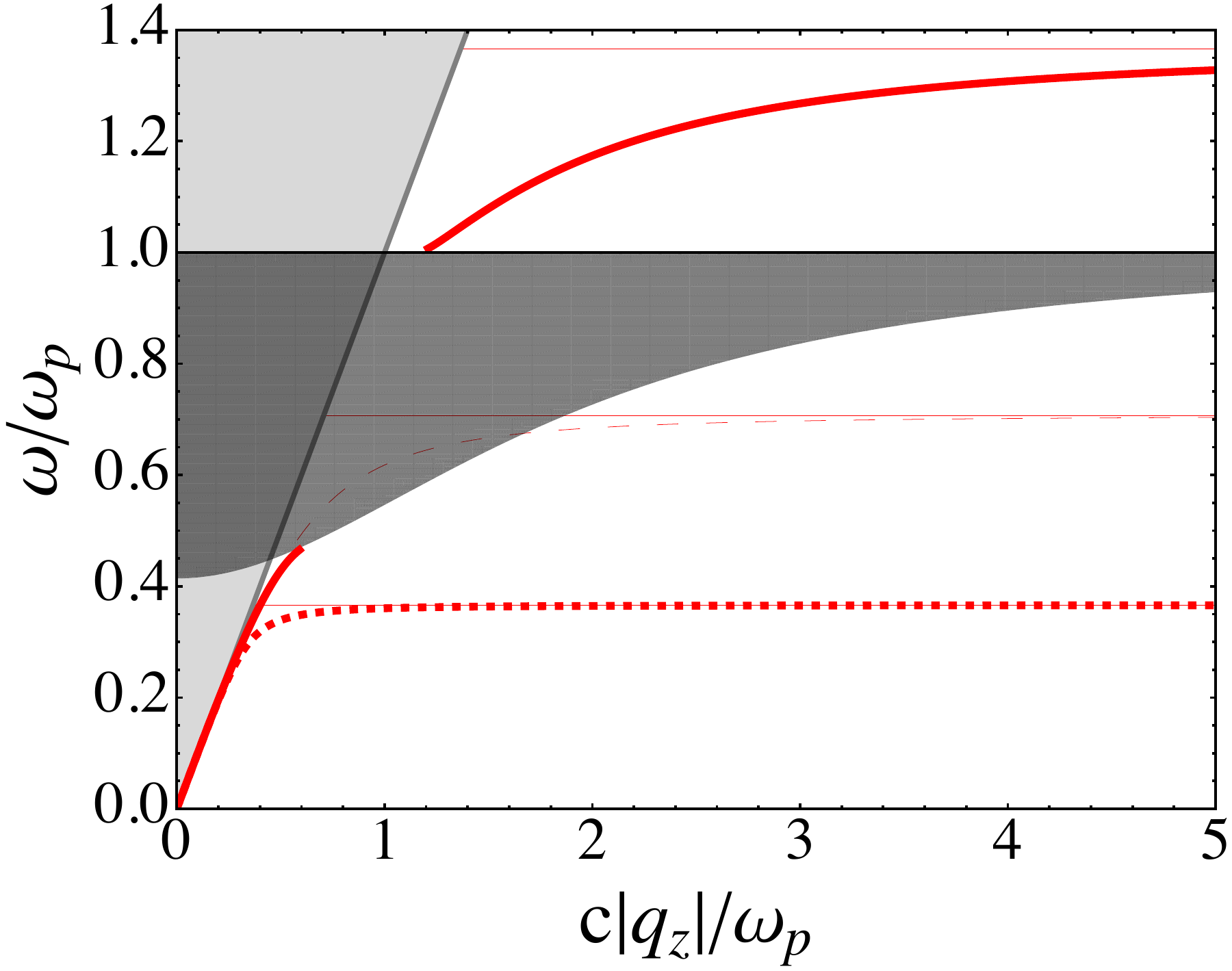}
    \caption{%
        \justifying Numerical SPP dispersion relations in the Voigt configuration with the AHE alone, for $ c \, G_{\mathrm{V}} / \omega_p = 2 $, plotted against $ | q_z | $. Thick solid red curve: $ q_z > 0 $; thick dotted red curve: $ q_z < 0 $; thin dashed red curve: conventional SPP. Thin horizontal red lines: asymptotes $ \omega_{\infty,s} ^{\mathrm{V}} $, Eq.~\eqref{eq:asymVoigt}, and $ \omega_\infty $, Eq.~\eqref{eq:omegainfty}. Light grey: dielectric light cone, Eq.~\eqref{eq:lightcone}; dark grey: forbidden region, Eq.~\eqref{eq:forbiddenregion}.
    }
    \label{fig:figure2}
\end{figure}

Unlike the conventional SPP frequency $ \omega_\infty $, which always lies below the plasma frequency, the upper asymptote, reached for $ \operatorname{sgn} \left( q_z \right) G_{\mathrm{V}} > 0 $, exceeds $ \omega_p $ when
\begin{equation}
    | G_{\mathrm{V}} | > \epsilon_d \frac{ \omega_p }{ c } .
\end{equation}
As observed in Ref.~\cite{OskouiAbdol2020}, a sufficiently strong AHE therefore allows SPPs to propagate above the plasma frequency, where $ \epsilon \left( \omega \right) > 0 $, but only in the direction with $ \operatorname{sgn} \left( q_z \right) = \operatorname{sgn} \left( G_{\mathrm{V}} \right) $, see Fig.~\ref{fig:figure2}.

Reversing the direction of propagation is equivalent to reversing $ \mathbf{G} $,
\begin{gather}
    \omega_{\mathrm{V}} \left( q_z, G_{\mathrm{V}} \right) = \omega_{\mathrm{V}} \left( - q_z, - G_{\mathrm{V}} \right).
\end{gather}

Since the bulk dispersion relation~\eqref{eq:bulk_imp_dispersion_relation} is even in $ \mathbf{q} $ for $ W = 0 $, this nonreciprocity arises from the boundary conditions: the full secular equation (see Appendix~\ref{app:PlanarVoigt}) contains terms odd in both $ q_z $ and $ G_{\mathrm{V}} $, which are invariant under their simultaneous reversal.

By contrast, in both the Faraday and perpendicular configurations, the large-wavevector expansion of the corresponding secular equations gives
\begin{equation}
    \omega_{\mathrm{F},\mathrm{P}} \left( q_z \to \pm \infty \right) = \omega_\infty - \frac{ \omega_\infty }{ 2 \left( \epsilon_\infty + \epsilon_d \right) } \left( \frac{ \omega_\infty ^2 }{ c ^2 } \epsilon_d ^2 + \frac{ G_{\mathrm{F},\mathrm{P}} ^2 }{ 4 } \right) \frac{ 1 }{ q_z ^2 } + \omega_p \, \mathcal{O} \left( \frac{ \omega_p ^4 }{ c ^4 q_z ^4 } \right).
\end{equation}

The vector $ \mathbf{G} $ leaves the asymptote unchanged and enters only the first correction, through $ G_{\mathrm{F},\mathrm{P}} ^2 $, which lowers the dispersion curve. The expansion depends on $ q_z $ only through $ q_z ^2 $, consistently with the reciprocity of the full dispersion relation in both configurations~\cite{Hofmann2016,Tamaya2019}, which is also even in $ \mathbf{G} $,
\begin{equation}
    \omega_{\mathrm{F},\mathrm{P}} \left( q_z , G_{\mathrm{F},\mathrm{P}} \right) = \omega_{\mathrm{F},\mathrm{P}} \left( q_z , - G_{\mathrm{F},\mathrm{P}} \right) .
\end{equation}

In all three configurations, the dispersion relations increase monotonically with $ | q_z | $ within their domain of existence,
\begin{equation}
    \operatorname{sgn} \left( q_z \right) \frac{ \mathrm{d} \omega_{\mathrm{V},\mathrm{F},\mathrm{P}} }{ \mathrm{d} q_z } > 0 .
\end{equation}
With the AHE alone, the group velocity is therefore always parallel to the wavevector, as shown in Fig.~\ref{fig:figure2} for the Voigt configuration. Consistently, the coefficients of $ 1 / q_z ^2 $ in the large-wavevector expansions are negative, and all branches approach their asymptotes from below.

Near the dielectric light line, moreover, the three dispersions behave in the same way. The expansion at small wavevector $ c | q_z | \ll \omega_p $ gives
\begin{equation}
    \omega_{\mathrm{V},\mathrm{F},\mathrm{P}} \left( q_z \to 0 \right) = \frac{ c | q_z | }{ \sqrt{ \epsilon_d } } - \frac{ c ^3 | q_z | ^3 }{ 2 \sqrt{ \epsilon_d } \epsilon_\infty \omega_p ^2 } + \omega_p \, \mathcal{O} \left( \frac{ c ^5 q_z ^5 }{ \omega_p ^5 } \right) ,
\end{equation}
which, at this order, does not depend on $ \mathbf{G} $ and coincides with that of a conventional SPP. In this regime, the mode is weakly confined and resides mostly in the dielectric, and the AHE enters only at higher orders (see Appendix~\ref{app:PlanarFaraday}).

\subsection{CME alone}
\label{sec:PlanarCME}

For $ W \ne 0 $ and $ \mathbf{G} = \mathbf{0} $ the conductor is isotropic, and the dispersion relation cannot depend on the direction of propagation along the interface. This case was considered numerically in Ref.~\cite{Hofmann2016} and, in a slot waveguide where the CME and the AHE act in two different conductors separated by a dielectric layer, in Ref.~\cite{OskouiAbdol2020}.

\begin{figure}[htbp]
    \centering
    \includegraphics[width=0.9\textwidth]{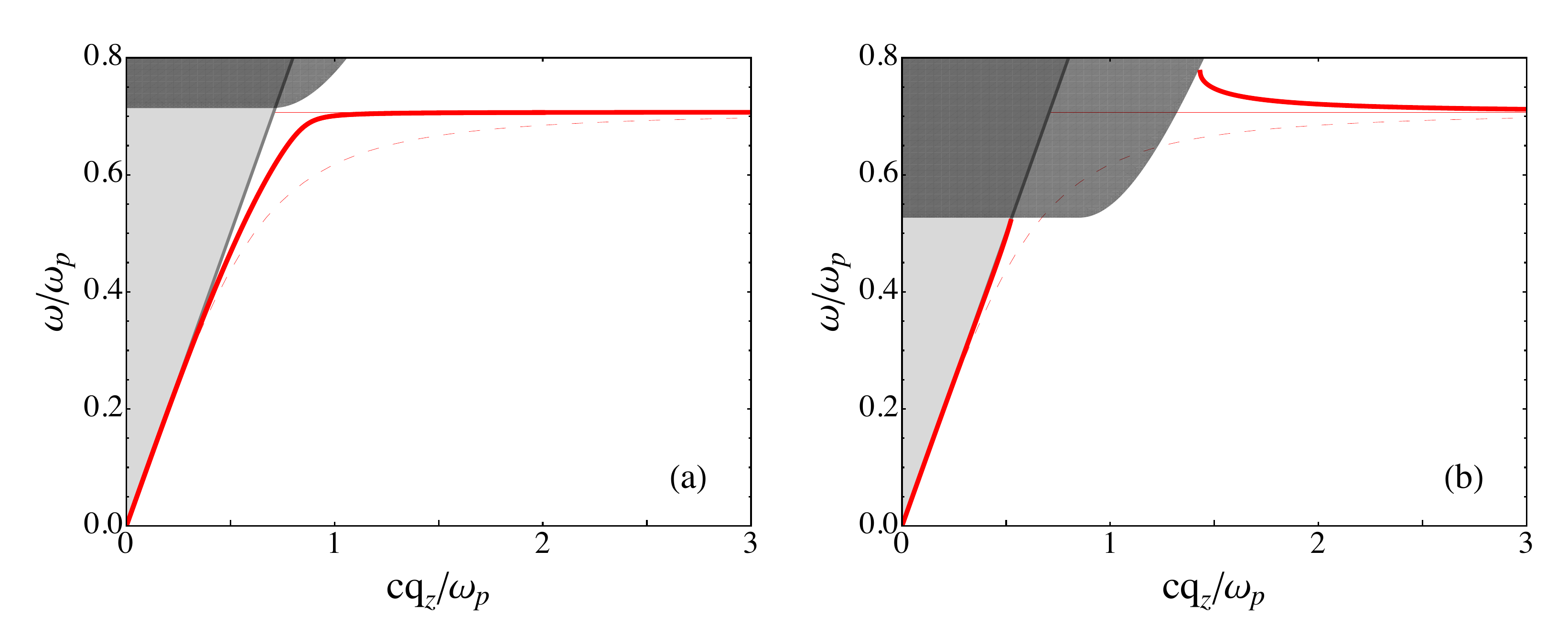}
    \caption{%
        \justifying Numerical SPP dispersion relation with the CME alone, for $ c \, W / \omega_p = 1.4 $ \textbf{(a)} and $ c \, W / \omega_p = 1.7 $ \textbf{(b)}, below and above the threshold $ c \, | W | / \omega_p = \sqrt{ 2 } $ for the reversal of the group velocity at large wavevector. Thick solid red curve: SPP, shown for $ q_z > 0 $ only since the dispersion is reciprocal; thin dashed red curve: conventional SPP. Thin horizontal red line: $ \omega_\infty $, Eq.~\eqref{eq:omegainfty}. Light grey: dielectric light cone, Eq.~\eqref{eq:lightcone}; dark grey: forbidden region, Eq.~\eqref{eq:forbiddenregion}.
    }
    \label{fig:figure3}
\end{figure}

Figure~\ref{fig:figure3} shows the dispersion relation for two values of $ W $, which display a qualitatively different behaviour at large wavevector, where the expansion of the secular equation gives
\begin{equation}\label{eq:PlanarVelocityCME}
    \omega_{\mathrm{CME}} \left( q_z \to \pm \infty \right) = \omega_\infty - \frac{ \epsilon_d ^2 \omega_\infty ^3 }{ 2 c ^2 \left( \epsilon_\infty + \epsilon_d \right) } \left( 1 - \frac{ c ^2 W ^2 }{ 4 \epsilon_d \, \omega_\infty ^2 } \right) \frac{ 1 }{ q_z ^2 } + \omega_p \, \mathcal{O} \left( \frac{ \omega_p ^4 }{ c ^4 q_z ^4 } \right).
\end{equation}

The asymptote is again $ \omega_\infty $, and $ W $ enters only the first correction, through $ W ^2 $, with the opposite sign with respect to $ \mathbf{G} $ in the Faraday and perpendicular configurations. As follows from the secular equation (see Appendix~\ref{app:PlanarFaraday}), the dispersion is reciprocal and even in $ W $, 
\begin{equation}
    \omega_{\mathrm{CME}} \left( q_z , W \right) = \omega_{\mathrm{CME}} \left( q_z , - W \right) .
\end{equation}

Nevertheless, the CME can change the shape of the dispersion qualitatively. The sign of the first correction in Eq.~\eqref{eq:PlanarVelocityCME} determines whether the dispersion approaches the asymptote from below or from above, and hence the direction of the group velocity at large wavevector,
\begin{align}
    \operatorname{sgn} \left( q_z \right) \left( \frac{ \mathrm{d} \omega_{\mathrm{CME}} }{ \mathrm{d} q_z } \right)_{ \displaystyle q_z \to \pm \infty } &\gtrless 0 , &
    \left( 1 - \frac{ c ^2 W ^2 }{ 4 \epsilon_d \, \omega_\infty ^2 } \right) &\gtrless 0 .
\end{align}
For $ | W | < 2 \sqrt{ \epsilon_d } \, \omega_\infty / c $, the dispersion relation increases monotonically with $ | q_z | $ within its domain of existence, as with the AHE alone; this is the case of Fig.~\ref{fig:figure3}(a). Above this threshold, the forbidden region splits it into two branches with opposite monotonicity: the one starting from the dielectric light line still increases with $ | q_z | $, while the one beyond the forbidden region decreases and approaches $ \omega_\infty $ from above, with a group velocity antiparallel to the wavevector, as shown in Fig.~\ref{fig:figure3}(b).

Near the light line, the expansion at small wavevector gives
\begin{equation}
    \omega_{\mathrm{CME}} \left( q_z \to 0 \right) = \frac{ c | q_z | }{ \sqrt{ \epsilon_d } } - \frac{ c ^3 | q_z | ^3 }{ 2 \sqrt{ \epsilon_d } \epsilon_\infty \omega_p ^2 } \left( 1 - \frac{ c ^2 W ^2 }{ 4 \epsilon_\infty \omega_p ^2 } \right) + \omega_p \, \mathcal{O} \left( \frac{ c ^5 q_z ^5 }{ \omega_p ^5 } \right) .
\end{equation}

Unlike $ \mathbf{G} $, the CME modifies the dispersion already at this order, reducing the cubic correction that keeps the curve below the light line. If the coefficient of $ | q_z | ^3 $ changed sign, the mode would enter the light cone of the dielectric at small wavevector and would no longer be bound to the interface. The propagation of SPPs therefore requires
\begin{equation}
    | W | < 2 \, \sqrt{ \epsilon_\infty } \, \frac{ \omega_p }{ c } ,
\end{equation}
which also holds in the presence of $ \mathbf{G} $. The bound involves only the permittivity parameters of the conductor, and not those of the dielectric (see Appendix~\ref{app:PlanarFaraday}).

\subsection{AHE in Voigt configuration with CME}
\label{sec:PlanarVoigt}

We now consider the coexistence of the two effects, starting with the Voigt configuration. In this case, the CME leaves the asymptotes of the AHE alone unchanged, and enters again only the first correction of the large-wavevector expansion, through $ W ^2 $,
\begin{equation}
    \omega_{\mathrm{VCME}} \left( q_z \to \pm \infty \right) = \omega_{\infty,s} ^{\mathrm{V}} - \frac{ \epsilon_d ^2 \left( \omega_{\infty,s} ^{\mathrm{V}} \right) ^4 }{ c ^2 \sqrt{ c ^2 G_{\mathrm{V}} ^2 + 4 \left( \epsilon_\infty + \epsilon_d \right) \epsilon_\infty \omega_p ^2 } } \left[ 1 - \frac{ c ^2 W ^2 }{ 4 \epsilon_d \left( \omega_{\infty,s} ^{\mathrm{V}} \right) ^2 } \right] \frac{ 1 }{ q_z ^2 } + \omega_p \, \mathcal{O} \left( \frac{ \omega_p ^4 }{ c ^4 q_z ^4 } \right) .
\end{equation}

\begin{figure}[htbp]
    \centering
    \includegraphics[width=0.9\textwidth]{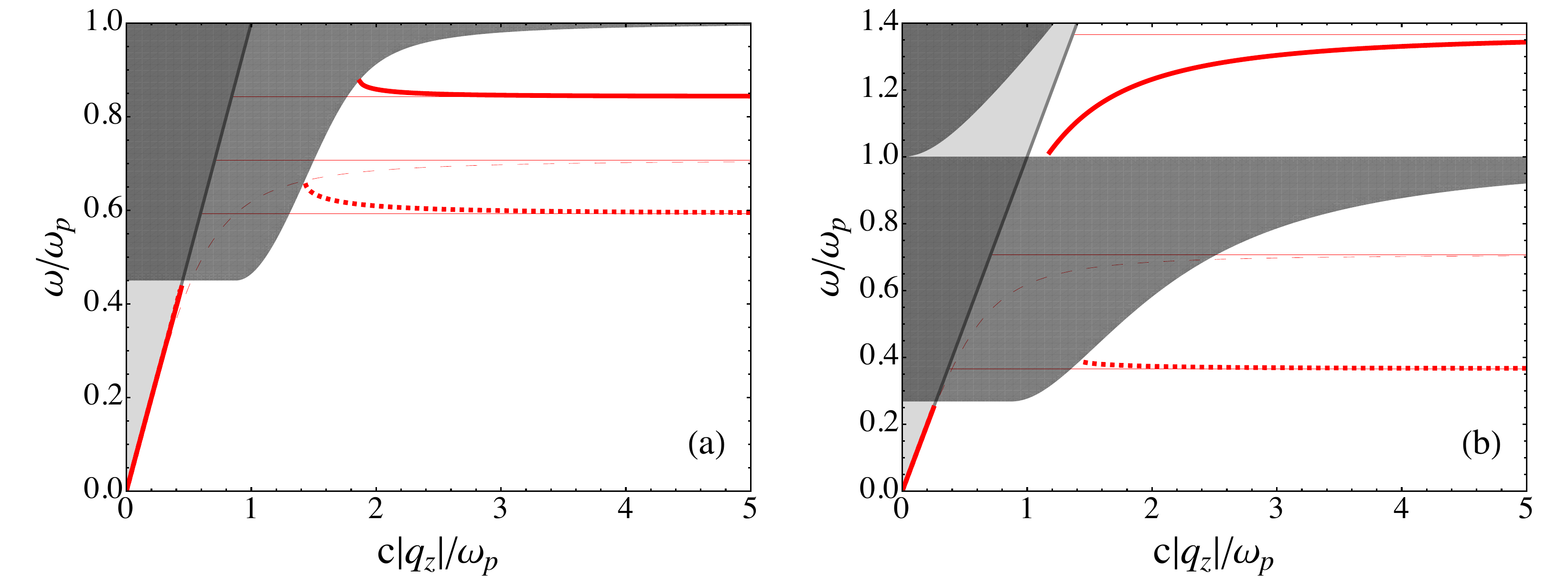}
    \caption{%
        \justifying Numerical SPP dispersion relations in the Voigt configuration with both effects, plotted against $ | q_z | $.
        Panel \textbf{(a)}: $ c \, G_{\mathrm{V}} / \omega_p = 0.5 $ and $ c \, W / \omega_p = 1.75 $; the group velocity at large wavevector is reversed for both directions.
        Panel \textbf{(b)}: $ c \, G_{\mathrm{V}} / \omega_p = 2 $ and $ c \, W / \omega_p = 1.75 $; it is reversed only for $ q_z < 0 $.
        Thick solid red curve: $ q_z > 0 $; thick dotted red curve: $ q_z < 0 $; thin dashed red curve: conventional SPP. Thin horizontal red lines: asymptotes $ \omega_{\infty,s} ^{\mathrm{V}} $, Eq.~\eqref{eq:asymVoigt}, and $ \omega_\infty $, Eq.~\eqref{eq:omegainfty}. Light grey: dielectric light cone, Eq.~\eqref{eq:lightcone}; dark grey: forbidden region, Eq.~\eqref{eq:forbiddenregion}.
    }
    \label{fig:figure4}
\end{figure}

The dispersion relation satisfies
\begin{equation}
    \omega_{\mathrm{VCME}} \left( q_z , G_{\mathrm{V}} , W \right) = \omega_{\mathrm{VCME}} \left( - q_z , - G_{\mathrm{V}} , W \right) = \omega_{\mathrm{VCME}} \left( q_z , G_{\mathrm{V}} , - W \right) .
\end{equation}
The CME does not add a nonreciprocity of its own, but it modifies the one produced by $ G_{\mathrm{V}} $, as shown in Fig.~\ref{fig:figure4}.

The corresponding group velocity at large wavevector is
\begin{equation}
    \left( \frac{ \mathrm{d} \omega_{\mathrm{VCME}} }{ \mathrm{d} q_z } \right)_{ \displaystyle q_z \to \pm \infty } = \frac{ 2 \epsilon_d ^2 \left( \omega_{\infty,s} ^{\mathrm{V}} \right) ^4 }{ c ^2 \sqrt{ c ^2 G_{\mathrm{V}} ^2 + 4 \left( \epsilon_\infty + \epsilon_d \right) \epsilon_\infty \omega_p ^2 } } \left[ 1 - \frac{ c ^2 W ^2 }{ 4 \epsilon_d \left( \omega_{\infty,s} ^{\mathrm{V}} \right) ^2 } \right] \frac{ 1 }{ q_z ^3 } + c \, \mathcal{O} \left( \frac{ \omega_p ^5 }{ c ^5 q_z ^5 } \right) .
\end{equation}

As for the CME alone, the sign of $ \operatorname{sgn} \left( q_z \right) \, \mathrm{d} \omega_{\mathrm{VCME}} / \mathrm{d} q_z $, which tells whether the group velocity is parallel or antiparallel to the wavevector, is controlled by the term in square brackets. This term now depends on the direction of propagation through $ \omega_{\infty,s} ^{\mathrm{V}} $, and the threshold for the reversal of the group velocity, $ | W | > 2 \sqrt{ \epsilon_d } \, \omega_{\infty,s} ^{\mathrm{V}} / c $, is different for the two branches. Again, only the part of each branch beyond the forbidden region is affected, while the one starting from the dielectric light line still increases with $ | q_z | $.

The threshold is reached first by the branch with the lower asymptote. For intermediate values of $ W $, the reversal affects one direction of propagation only, and the group velocities of the two branches have the same sign, fixed by that of $ G_{\mathrm{V}} $. Both directions are reversed only if the threshold of the upper branch lies below the bound on $ W $, that is, $ \omega_{\infty,s} ^{\mathrm{V}} < \sqrt{ \epsilon_\infty / \epsilon_d } \, \omega_p $, as in Fig.~\ref{fig:figure4}(a). For a strong AHE, the upper asymptote exceeds this value and only one direction can be reversed, while the propagation above $ \omega_p $ persists, since $ W $ leaves the asymptotes unchanged: this case is illustrated in Fig.~\ref{fig:figure4}(b).

\subsection{AHE in Faraday configuration with CME}
\label{sec:PlanarFaraday}

\begin{figure}[ht]
    \centering
    \includegraphics[width=0.9\textwidth]{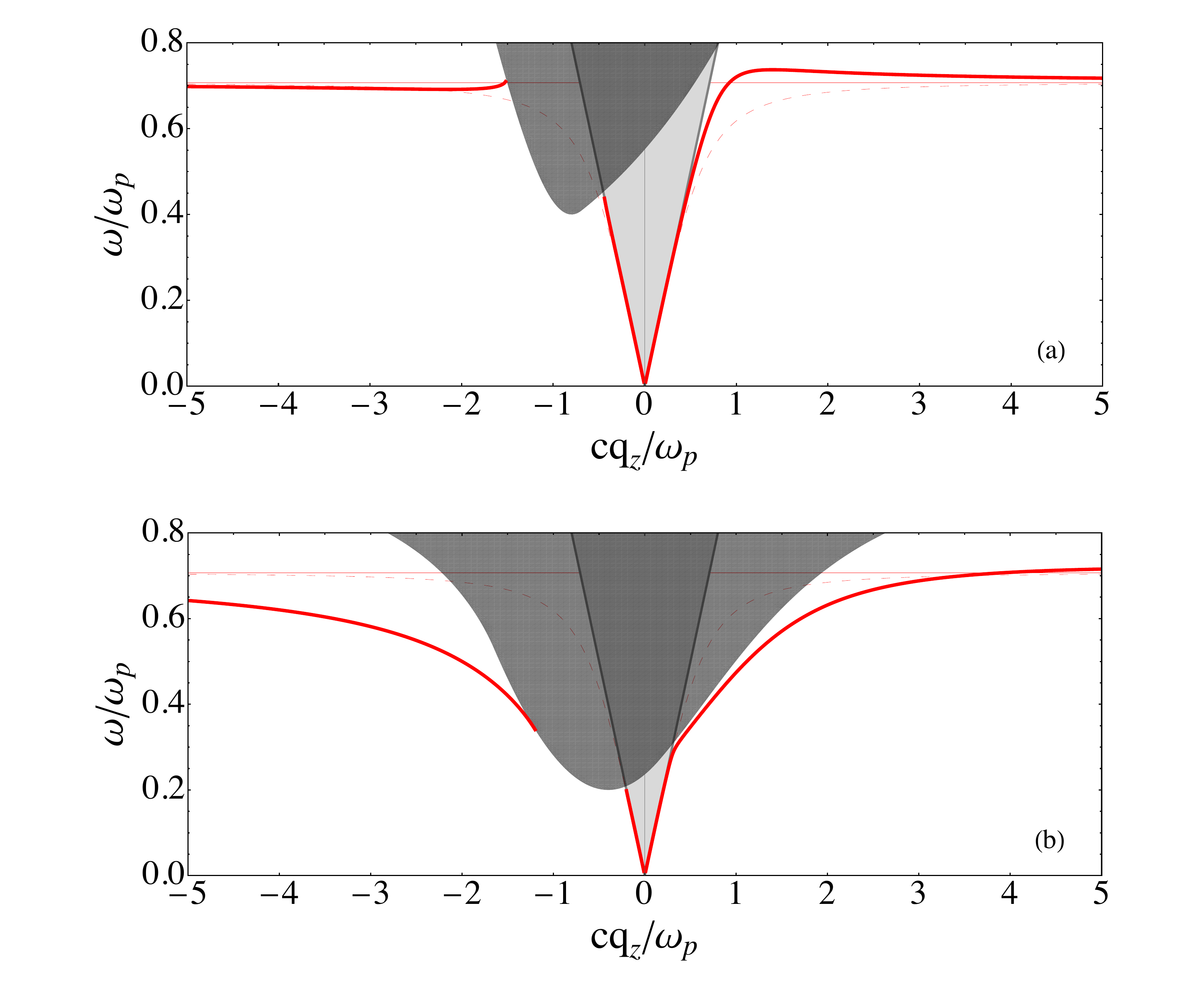}
    \caption{%
        \justifying Numerical SPP dispersion relations in the Faraday configuration with both effects, for both signs of $ q_z $.
        Panel \textbf{(a)}: $ c \, G_{\mathrm{F}} / \omega_p = 0.5 $ and $ c \, W / \omega_p = 1.6 $; the branch with $ q_z > 0 $ has a maximum and the one with $ q_z < 0 $ a minimum.
        Panel \textbf{(b)}: $ c \, G_{\mathrm{F}} / \omega_p = 4 $ and $ c \, W / \omega_p = 0.8 $; the two branches approach $ \omega_\infty $ from opposite sides.
        Thick solid red curves: SPP; thin dashed red curve: conventional SPP. Thin horizontal red line: $ \omega_\infty $, Eq.~\eqref{eq:omegainfty}. Light grey: dielectric light cone, Eq.~\eqref{eq:lightcone}; dark grey: forbidden region, Eq.~\eqref{eq:forbiddenregion}.
    }
    \label{fig:figure5}
\end{figure}

In the Faraday configuration, the coexistence of the two effects changes the structure of the large-wavevector expansion, as a term of order $ 1 / q_z $ appears, proportional to the product $ G_{\mathrm{F}} W $,
\begin{multline}
    \omega_{\mathrm{FCME}} \left( q_z \to \pm \infty \right) = \omega_\infty + \frac{ c \, G_{\mathrm{F}} W }{ 8 \left( \epsilon_\infty + \epsilon_d \right) } \frac{ 1 }{ q_z } \\
    - \frac{ \epsilon_d ^2 \, \omega_\infty ^3 }{ 2 c ^2 \left( \epsilon_\infty + \epsilon_d \right) } \left[ 1 + \frac{ c ^2 G_{\mathrm{F}} ^2 }{ 4 \epsilon_d ^2 \omega_\infty ^2 } - \frac{ c ^2 W ^2 }{ 4 \epsilon_d \, \omega_\infty ^2 } - \frac{ c ^4 G_{\mathrm{F}} ^2 W ^2 }{ 64 \, \epsilon_d ^2 \left( \epsilon_\infty + \epsilon_d \right) \omega_\infty ^4 } \right] \frac{ 1 }{ q_z ^2 } + \omega_p \, \mathcal{O} \left( \frac{ \omega_p ^3 }{ c ^3 q_z ^3 } \right) .
\end{multline}
The asymptote is still $ \omega_\infty $, but the new term is odd in $ q_z $, and the two directions of propagation approach it differently, as shown in Fig.~\ref{fig:figure5}. The dispersion relation satisfies
\begin{equation}
    \omega_{\mathrm{FCME}} \left( q_z , G_{\mathrm{F}} , W \right) = \omega_{\mathrm{FCME}} \left( - q_z , - G_{\mathrm{F}} , W \right) = \omega_{\mathrm{FCME}} \left( - q_z , G_{\mathrm{F}} , - W \right) ,
\end{equation}
that is, reversing the direction of propagation is equivalent to reversing either $ G_{\mathrm{F}} $ or $ W $. Neither effect alone breaks reciprocity in this configuration, but their product does, as a consequence of the nonzero bulk term $ W \left( \mathbf{q} \cdot \mathbf{G} \right) = W G_{\mathrm{F}} q_z $, odd in $ q_z $. Unlike the Voigt configuration, where the nonreciprocity already appears in the asymptotes, here it appears only in the first correction.

At large wavevector, the group velocity is dominated by the term in $ G_{\mathrm{F}} W $,
\begin{equation}
    \left( \frac{ \mathrm{d} \omega_{\mathrm{FCME}} }{ \mathrm{d} q_z } \right)_{ \displaystyle q_z \to \pm \infty } = - \frac{ c \, G_{\mathrm{F}} W }{ 8 \left( \epsilon_\infty + \epsilon_d \right) } \frac{ 1 }{ q_z ^2 } + c \, \mathcal{O} \left( \frac{ \omega_p ^3 }{ c ^3 q_z ^3 } \right) ,
\end{equation}
and has the same sign, opposite to that of $ G_{\mathrm{F}} W $, for both directions of propagation. One branch therefore approaches $ \omega_\infty $ from above and the other from below, as displayed in Fig.~\ref{fig:figure5}(b).

Unlike the previous cases, the dispersion relation is no longer monotonic even within each part of its domain of existence. Since the term in $ 1 / q_z $ changes sign with the direction of propagation while the one in $ 1 / q_z ^2 $ does not, the two corrections compete on one of the two branches, and the dispersion relation can reach extrema where the group velocity vanishes, as shown in Fig.~\ref{fig:figure5}(a), in which the branch with $ q_z > 0 $ has a maximum and the one with $ q_z < 0 $ a minimum.

\subsection{AHE in perpendicular configuration with CME}
\label{sec:PlanarPerp}

When $ \mathbf{G} $ is normal to the interface, the coupling term $ W \left( \mathbf{q} \cdot \mathbf{G} \right) $ involves the normal component of the wavevector, which is imaginary for the evanescent waves in the conductor,
\begin{equation}
    W \left( \mathbf{q} \cdot \mathbf{G} \right) = - \mathrm{i} \kappa_\pm G_{\mathrm{P}} W,
\end{equation}
which implies that the bulk dispersion relation acquires an imaginary coefficient (see Appendix~\ref{app:PlanarPerpendicular}). To discuss its consequences, we consider the transformations
\begin{align}
    \mathfrak{T}: \left( q_z , \omega , \mathbf{G} , W \right) &\to \left( - q_z , \omega ^\star , - \mathbf{G} , W \right) , \\
    \mathfrak{R}_{y,\pi}: \left( q_z , \mathbf{G}_\perp ^{ \left( y \right) } \right) &\to \left( - q_z, - \mathbf{G}_\perp ^{ \left( y \right) } \right) , \\
    \mathfrak{M}_z: \left( z , q_z , \mathbf{G}_\perp ^{ \left( z \right) } , W \right) &\to \left( - z , - q_z, - \mathbf{G}_\perp ^{ \left( z \right) } , - W \right) ,
\end{align}
where $ \mathfrak{T} $ is time reversal, which in the frequency domain maps $ \omega $ into its complex conjugate $ \omega ^\star $~\cite{Caloz2018}, $ \mathfrak{R}_{y,\pi} $ is the rotation by an angle $ \pi $ about the normal to the interface, $ \mathfrak{M}_z $ is the reflection through the plane $ z = 0 $, and $ \mathbf{G}_\perp ^{ \left( y \right) } $ and $ \mathbf{G}_\perp ^{ \left( z \right) } $ denote the components of $ \mathbf{G} $ orthogonal to $ y $ and to $ z $, respectively. Each of these transformations reverses $ q_z $, while their combinations with time reversal, as in the magnetic symmetry groups of crystals~\cite{Landau1984}, leave it unchanged,
\begin{align}
    \mathfrak{T} \, \mathfrak{R}_{y,\pi} &: \left( \mathbf{G} \cdot \mathbf{e}_y , W \right) \to \left( - \mathbf{G} \cdot \mathbf{e}_y , W \right) , \\
    \mathfrak{T} \, \mathfrak{M}_z &: \left( \mathbf{G} \cdot \mathbf{e}_z , W \right) \to \left( - \mathbf{G} \cdot \mathbf{e}_z , - W \right) ,
\end{align}
with the other components of $ \mathbf{G} $ unchanged. Thus, $ \mathfrak{T} \, \mathfrak{R}_{y,\pi} $ reverses only the normal component of $ \mathbf{G} $, while $ \mathfrak{T} \, \mathfrak{M}_z $ reverses its component along $ z $ and $ W $. Since both map a solution $ \omega $ into $ \omega ^\star $, whenever one of them is a symmetry of the system the secular equation satisfies $ \det{ \mathcal{M} \left( \omega ^\star \right) } = \left[ \det{ \mathcal{M} \left( \omega \right) } \right] ^\star $, and its complex solutions come in conjugate pairs, as for $ \mathcal{PT} $-symmetric systems~\cite{Bender1998}. Real solutions, which describe the freely propagating SPPs we are interested in, are then protected: a small perturbation that preserves the symmetry cannot move a simple real solution off the real axis. This is the case in all the configurations discussed so far, in which at least one of the two symmetries is preserved. In the perpendicular configuration with the CME, instead, $ \mathfrak{T} \, \mathfrak{R}_{y,\pi} $ is broken by $ G_{\mathrm{P}} $ and $ \mathfrak{T} \, \mathfrak{M}_z $ by $ W $. Then
\begin{equation}
    \det{ \mathcal{M}_{\mathrm{PCME}} \left( \omega ^\star \right) } \neq \left[ \det{ \mathcal{M}_{\mathrm{PCME}} \left( \omega \right) } \right] ^\star ,
\end{equation}
and nothing protects the real solutions. The expansion of the secular equation at large wavevector confirms that they are lost,
\begin{equation}
    \epsilon \left( \omega \right) + \epsilon_d = - \frac{ \mathrm{i} \, c \, G_{\mathrm{P}} W }{ 4 \omega } \frac{ 1 }{ | q_z | } - \left( \frac{ \omega ^2 }{ c ^2 } \epsilon_d ^2 + \frac{ G_{\mathrm{P}} ^2 }{ 4 } - \frac{ \epsilon_d W ^2 }{ 4 } \right) \frac{ 1 }{ q_z ^2 } + \mathcal{O} \left( \frac{ \omega_p ^3 }{ c ^3 q_z ^3 } \right) .
\end{equation}
The term of order $ 1 / | q_z | $ is purely imaginary and proportional to $ G_{\mathrm{P}} W $, and gives $ \omega $ an imaginary part, while the real terms reproduce the structure found for the AHE and the CME alone (see Appendix~\ref{app:PlanarPerpendicular}). Therefore, when $ W \neq 0 $ and $ \mathbf{G} $ is normal to the interface, no freely propagating surface modes exist, however small the product $ G_{\mathrm{P}} W $; real solutions are recovered as soon as either $ G_{\mathrm{P}} $ or $ W $ vanishes.

\section{Cylindrical Interface}
\label{sec:Cylindrical}

We now turn to the cylindrical interface of Fig.~\ref{fig:figure1}(b), with $ \mathbf{G} = G_z \mathbf{e}_z $ along the axis. Compared with the planar geometry, the curvature of the interface introduces the radius $ R $ as a new length scale, and the modes are labelled not only by the wavevector $ q_z $, which we also refer to as the linear momentum, but also by the OAM $ m $, which counts the windings of the phase around the axis. The secular equation for each $ m $ is derived in Appendix~\ref{app:Cylindrical}. We expand the dispersion relations at large $ | q_z | $ for fixed $ m $, where the corrections due to the curvature appear as powers of $ 1 / \left( R q_z \right) $, and at large $ | m | $ for fixed $ q_z $, where the mode is dominated by its azimuthal structure.

\subsection{AHE alone}
\label{sec:CylAHE}

For $ \mathbf{G} \ne \mathbf{0} $ and $ W = 0 $ we recover the results of Ref.~\cite{Peluso2025WG}, which we express here in the form of asymptotic expansions.

At large linear momentum, the dispersion relation reads
\begin{align}
    \omega_m ^{\mathrm{AHE}} \left( q_z \to \pm \infty \right) = \omega_\infty &+ \frac{ c \, G_z m - \epsilon_d \omega_\infty }{ 2 \left( \epsilon_\infty + \epsilon_d \right) } \frac{ 1 }{ R | q_z | } - \frac{ \epsilon_d ^2 \, \omega_\infty ^3 }{ 2 c ^2 \left( \epsilon_\infty + \epsilon_d \right) } \left( 1 + \frac{ c ^2 G_z ^2 }{ 4 \epsilon_d ^2 \omega_\infty ^2 } \right) \frac{ 1 }{ q_z ^2 } \nonumber \\[8pt]
    &+ \frac{ \left( c \, G_z m - \epsilon_d \omega_\infty \right) \left( c \, G_z m - \epsilon_d \omega_\infty + 2 \epsilon_\infty \omega_\infty \right) }{ 8 \, \omega_\infty \left( \epsilon_\infty + \epsilon_d \right) ^2 } \frac{ 1 }{ R ^2 q_z ^2 } + \omega_p \, \mathcal{O} \left( \frac{ \omega_p ^3 }{ c ^3 q_z ^3 } \right) .
\end{align}

The asymptote is the planar one, $ \omega_\infty $, as expected when the wavelength is much shorter than the radius. The first correction, of order $ 1 / \left( R | q_z | \right) $, contains the curvature term $ - \epsilon_d \omega_\infty $, already present for a conventional conductor, and the term $ c \, G_z m $, which is linear in both $ G_z $ and $ m $ and lifts the degeneracy between modes with opposite OAM. The term in $ G_z ^2 / q_z ^2 $ coincides with that of the planar Faraday configuration, in which $ \mathbf{G} $ is likewise parallel to the wavevector. The expansion depends on $ q_z $ only through $ | q_z | $, consistently with the secular equation of Appendix~\ref{app:CylindricalCond}: the propagation along the axis is reciprocal.

\begin{figure}[htbp]
    \centering
    \includegraphics[width=0.45\textwidth]{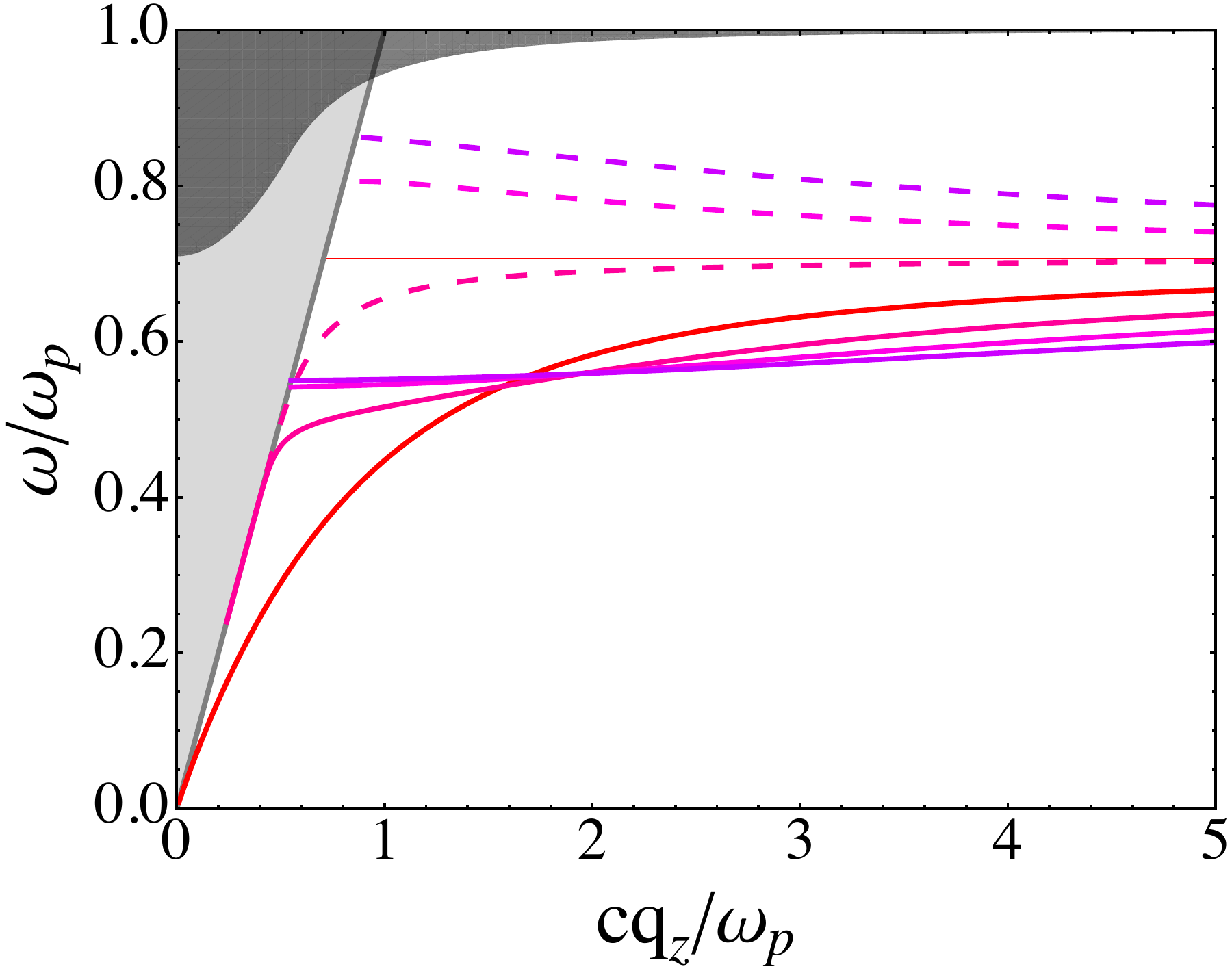}
    \caption{%
        \justifying Numerical SPP dispersion relations at the cylindrical interface with the AHE alone, for $ \omega_p R / c = 1 $ and $ c \, G_z / \omega_p = 0.7 $. The propagation along the axis is reciprocal, and only $ q_z > 0 $ is shown. Solid curves: $ m \leq 0 $; dashed curves: $ m > 0 $. The $ m = 0 $ mode is drawn in red, while the modes with $ \left| m \right| = 1 , 2 , 3 $ are drawn in magenta, fuchsia and violet, respectively. Thin horizontal red line: $ \omega_\infty $, Eq.~\eqref{eq:omegainfty}. Thin horizontal dark violet lines: large-OAM asymptotes $ \omega_{\infty,m} ^{\mathrm{AHE}} $, Eq.~\eqref{eq:asymAHEm}, solid for $ m < 0 $ and dashed for $ m > 0 $. Light grey: dielectric light cone, Eq.~\eqref{eq:lightcone}; dark grey: forbidden region, Eq.~\eqref{eq:forbiddenregion}.
    }
    \label{fig:figure6}
\end{figure}

At large OAM, the expansion of the secular equation gives
\begin{multline}
    \omega_m ^{\mathrm{AHE}} \left( m \to \pm \infty \right) = \omega_{\infty,m} ^{\mathrm{AHE}} - \frac{ c \, \omega_{\infty,m} ^{\mathrm{AHE}} }{ 4 \sqrt{ c ^2 G_z ^2 + 4 \left( \epsilon_\infty + \epsilon_d \right) \epsilon_\infty \omega_p ^2 } } \\[8pt]
    \times \left\{ 2 \operatorname{sgn} \left( m \right) \, G_z q_z ^2 + \frac{ 4 \epsilon_d ^2 \, \left( \omega_{\infty,m} ^{\mathrm{AHE}} \right) ^3 }{ c ^3 } \right\} \frac{ R ^2 }{ m ^2 } + \omega_p \, \mathcal{O} \left( \frac{ 1 }{ m ^3 } \right) ,
\end{multline}
where
\begin{equation}\label{eq:asymAHEm}
  \omega_{\infty,m} ^{\mathrm{AHE}} \left[ \operatorname{sgn} \left( m \right) , G_z \right] = \frac{ \sqrt{ c ^2 G_z ^2 + 4 \left( \epsilon_\infty + \epsilon_d \right) \epsilon_\infty \omega_p ^2 } + \operatorname{sgn} \left( m \right) \, c \, G_z }{ 2 \left( \epsilon_\infty + \epsilon_d \right) } .
\end{equation}

In this regime the asymptote itself depends on the sign of $ m $, and has the same form as $ \omega_{\infty,s} ^{\mathrm{V}} $, with $ \operatorname{sgn} \left( m \right) $ in place of $ \operatorname{sgn} \left( q_z \right) $. For a mode dominated by its azimuthal structure, $ \mathbf{G} $ is tangent to the interface and orthogonal to the direction of the phase winding, and the two senses of rotation about the axis become inequivalent. Consequently, the modes with $ \operatorname{sgn} \left( m \right) G_z > 0 $ can propagate above $ \omega_p $ under the same condition as in the Voigt configuration, $ | G_z | > \epsilon_d \, \omega_p / c $.

The dispersion relation satisfies
\begin{equation}
    \omega_m ^{\mathrm{AHE}} \left( q_z, G_z \right) = \omega_{-m} ^{\mathrm{AHE}} \left( q_z, - G_z \right) ,
\end{equation}
so that, unlike the propagation along the axis, the AHE alone breaks reciprocity in $ m $.

Two further features of this system, discussed in Ref.~\cite{Peluso2025WG} and illustrated in Fig.~\ref{fig:figure6}, follow from the two expansions. The modes with $ \operatorname{sgn} \left( m \right) G_z < 0 $ are ordered differently at large $ | m | $, 
and their dispersion curves therefore cross each other: in a finite interval of $ q_z $, the mode $ m = 0 $ is no longer the lowest in frequency. Moreover, at large linear momentum the modes with $ c \, G_z m > \epsilon_d \, \omega_\infty $ approach $ \omega_\infty $ from above, so that modes with opposite $ m $ can propagate with opposite group velocities.

\subsection{CME alone}
\label{sec:CylCME}

\begin{figure}[htbp]
    \centering
    \includegraphics[width=0.9\textwidth]{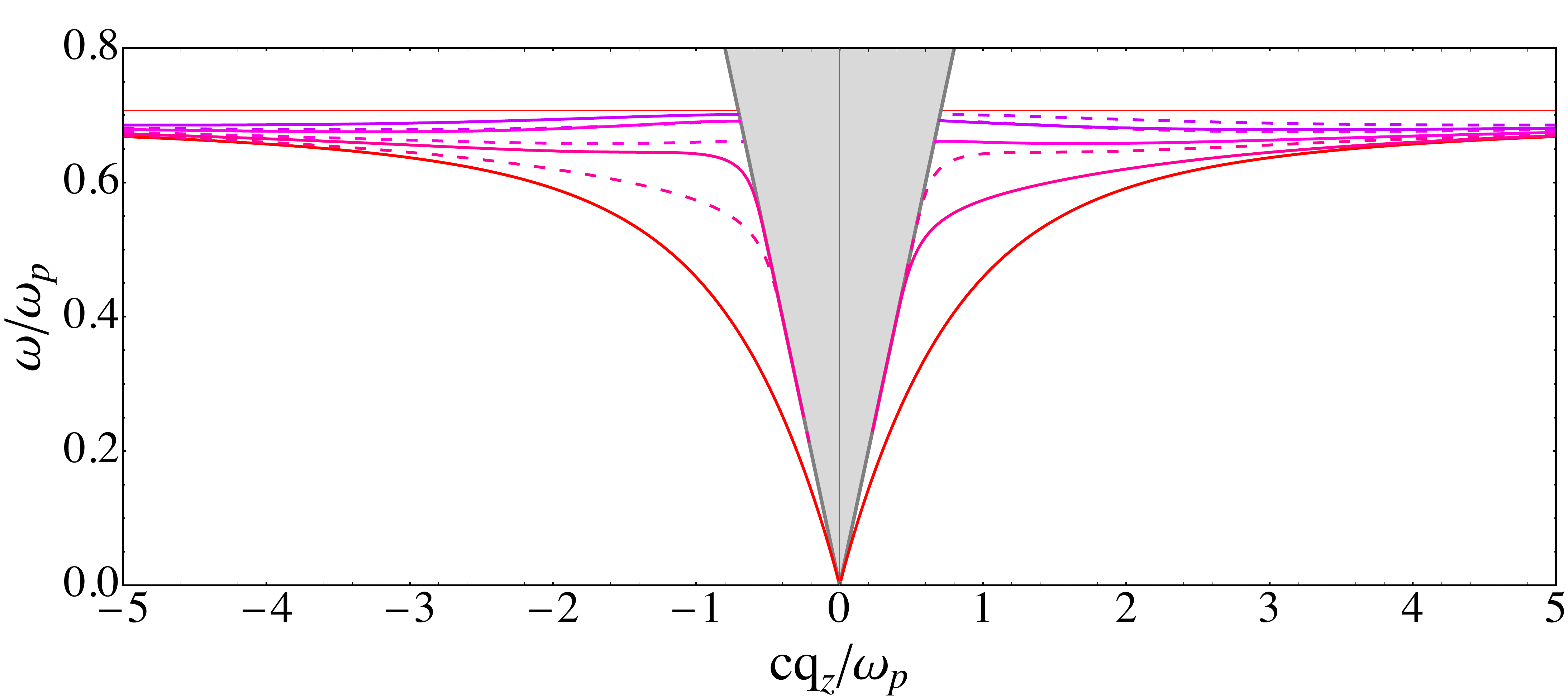}
    \caption{%
        \justifying Numerical SPP dispersion relations at the cylindrical interface with the CME alone, for $ \omega_p R / c = 1 $ and $ c \, W / \omega_p = 1 $, for both signs of $ q_z $. Solid curves: $ m \leq 0 $; dashed curves: $ m > 0 $. The $ m = 0 $ mode is drawn in red, while the modes with $ \left| m \right| = 1 , 2 , 3 $ are drawn in magenta, fuchsia and violet, respectively. Thin horizontal red line: $ \omega_\infty $, Eq.~\eqref{eq:omegainfty}. Light grey: dielectric light cone, Eq.~\eqref{eq:lightcone}; the forbidden region lies outside the frequency range shown.
    }
    \label{fig:figure7}
\end{figure}

We now set $ \mathbf{G} = \mathbf{0} $ and $ W \ne 0 $, for which the conductor is isotropic. Figure~\ref{fig:figure7} displays the numerical solutions of the secular equation: unlike the AHE, which acts on $ m $ only, the CME couples the OAM of the modes to their linear momentum.

The dispersion relation satisfies
\begin{equation}
    \omega_m ^{\mathrm{CME}} \left( q_z, W \right) = \omega_{-m} ^{\mathrm{CME}} \left( q_z, - W \right) = \omega_m ^{\mathrm{CME}} \left( - q_z, - W \right) .
\end{equation}

The sign of $ W $ fixes which of the two modes $ \pm m $ lies higher in frequency for a given direction of propagation. Reciprocity is broken in $ q_z $ and in $ m $ separately, and only their relative sign matters. Any combination odd in both $ m $ and $ q_z $ has the same transformation properties as $ W $, being even under time reversal and odd under parity: a correction linear in $ W $ must therefore involve such a combination.

The expansion of the dispersion relation at large linear momentum makes the coupling explicit,
\begin{align}
    \omega_m ^{\mathrm{CME}} \left( q_z \to \pm \infty \right) = \omega_\infty &- \frac{ \epsilon_d \omega_\infty }{ 2 \left( \epsilon_\infty + \epsilon_d \right) } \frac{ 1 }{ R | q_z | } - \frac{ \epsilon_d ^2 \, \omega_\infty ^3 }{ 2 c ^2 \left( \epsilon_\infty + \epsilon_d \right) } \left( 1 - \frac{ c ^2 W ^2 }{ 4 \epsilon_d \, \omega_\infty ^2 } \right) \frac{ 1 }{ q_z ^2 } \nonumber \\[8pt]
    &+ \frac{ \epsilon_d \left( \epsilon_d - 2 \epsilon_\infty \right) \omega_\infty }{ 8 \left( \epsilon_\infty + \epsilon_d \right) ^2 } \frac{ 1 }{ R ^2 q_z ^2 } \nonumber \\[8pt]
    &+ \frac{ \epsilon_d \omega_\infty \left[ 4 \epsilon_d \omega_\infty ^2 \left( 2 \epsilon_d - 3 \epsilon_\infty \right) - c ^2 W ^2 \left( \epsilon_d - 2 \epsilon_\infty \right) \right] }{ 16 \, c ^2 \left( \epsilon_\infty + \epsilon_d \right) ^2 } \frac{ 1 }{ R | q_z | ^3 } + \frac{ \epsilon_d W m \, \omega_\infty }{ 2 \left( \epsilon_\infty + \epsilon_d \right) } \frac{ 1 }{ R ^2 q_z ^3 } \nonumber \\[8pt]
    &+ \frac{ \epsilon_d \omega_\infty \left[ 12 m ^2 \left( \epsilon_\infty + \epsilon_d \right) ^2 - 4 \epsilon_d ^2 - 4 \epsilon_d \epsilon_\infty - 5 \epsilon_\infty ^2 \right] }{ 16 \left( \epsilon_\infty + \epsilon_d \right) ^3 } \frac{ 1 }{ R ^3 | q_z | ^3 } + \omega_p \, \mathcal{O} \left( \frac{ \omega_p ^4 }{ c ^4 q_z ^4 } \right) .
\end{align}
All the terms are even in $ W $ and depend on $ q_z $ only through its absolute value, except the one of order $ 1 / \left( R ^2 q_z ^3 \right) $, whose coefficient is proportional to $ W m $. This correction makes these symmetries explicit: it requires a nonzero OAM, and has no counterpart at the planar interface.

Since the leading correction, of order $ 1 / \left( R | q_z | \right) $,  is negative and independent of $ W $, all the modes eventually approach $ \omega_\infty $ from below.

At large OAM and fixed linear momentum, the dispersion relation reads 
\begin{equation}
    \omega_m ^{\mathrm{CME}} \left( m \to \pm \infty \right) = \omega_\infty - \frac{ \epsilon_d ^2 \, \omega_\infty ^3 }{ 2 c ^2 \left( \epsilon_\infty + \epsilon_d \right) } \left( 1 - \frac{ c ^2 W ^2 }{ 4 \epsilon_d \omega_\infty ^2 } \right) \frac{ R ^2 }{ m ^2 } + \omega_p \, \mathcal{O} \left( \frac{ 1 }{ m ^3 } \right) .
\end{equation}
In this regime, where the numerical solutions of the secular equation become demanding, the behaviour of the modes is directly accessible through the corresponding expansion, which shows that they approach $ \omega_\infty $ from above as $ | m | $ increases when $ | W | > 2 \sqrt{ \epsilon_d } \, \omega_\infty / c $.

\subsection{Interplay between AHE and CME}
\label{sec:CylBoth}

When both effects are present, the expansion at large linear momentum also contains terms proportional to $ G_z W $,
\begin{align}
    \omega_m ^{\mathrm{AHECME}} \left( q_z \to \pm \infty \right) = \omega_\infty &+ \frac{ c \, G_z W }{ 8 \left( \epsilon_\infty + \epsilon_d \right) } \frac{ 1 }{ q_z } + \frac{ c \, G_z m - \epsilon_d \omega_\infty }{ 2 \left( \epsilon_\infty + \epsilon_d \right) } \frac{ 1 }{ R | q_z | } \nonumber \\[8pt]
    &- \frac{ \epsilon_d ^2 \, \omega_\infty ^3 }{ 2 c ^2 \left( \epsilon_\infty + \epsilon_d \right) } \left[ 1 + \frac{ c ^2 G_z ^2 }{ 4 \epsilon_d ^2 \omega_\infty ^2 } - \frac{ c ^2 W ^2 }{ 4 \epsilon_d \, \omega_\infty ^2 } - \frac{ c ^4 G_z ^2 W ^2 }{ 64 \, \epsilon_d ^2 \left( \epsilon_\infty + \epsilon_d \right) \omega_\infty ^4 } \right] \frac{ 1 }{ q_z ^2 } \nonumber \\[8pt]
    &+ \frac{ \operatorname{sgn} \left( q_z \right) \, c \, G_z W \left( c \, G_z m - 2 \epsilon_d \omega_\infty \right) }{ 16 \, \omega_\infty \left( \epsilon_\infty + \epsilon_d \right) ^2 } \frac{ 1 }{ R q_z ^2 } \nonumber \\[8pt]
    &+ \frac{ \left( c \, G_z m - \epsilon_d \omega_\infty \right) \left( c \, G_z m - \epsilon_d \omega_\infty + 2 \epsilon_\infty \omega_\infty \right) }{ 8 \, \omega_\infty \left( \epsilon_\infty + \epsilon_d \right) ^2 } \frac{ 1 }{ R ^2 q_z ^2 } + \omega_p \, \mathcal{O} \left( \frac{ \omega_p ^3 }{ c ^3 q_z ^3 } \right) .
\end{align}
At first order in $ 1 / q_z $, the correction in $ G_z W $, already found in the planar Faraday configuration, is odd in $ q_z $ and independent of $ m $ and $ R $, while the one of order $ 1 / \left( R | q_z | \right) $ splits the modes of opposite OAM, as for the AHE alone. The nonreciprocity along the axis thus appears at order $ 1 / q_z $, rather than $ 1 / q_z ^3 $ as for the CME alone. In particular, the fundamental mode $ m = 0 $, which is reciprocal for either effect alone, becomes nonreciprocal when both are present. A mixed contribution, odd in $ q_z $ and involving $ R $, $ m $ and $ G_z W $ together, follows at order $ 1 / \left( R q_z ^2 \right) $.

At large OAM, the expansion gives
\begin{multline}
    \omega_m ^{\mathrm{AHECME}} \left( m \to \pm \infty \right) = \omega_{\infty,m} ^{\mathrm{AHE}} - \frac{ c \, \omega_{\infty,m} ^{\mathrm{AHE}} }{ 4 \sqrt{ c ^2 G_z ^2 + 4 \left( \epsilon_\infty + \epsilon_d \right) \epsilon_\infty \omega_p ^2 } } \\[8pt]
    \times \left\{ 2 \operatorname{sgn} \left( m \right) \, G_z q_z ^2 - G_z W q_z + \frac{ 4 \epsilon_d ^2 \left( \omega_{\infty,m} ^{\mathrm{AHE}} \right) ^3 }{ c ^3 } \left[ 1 - \frac{ c ^2 W ^2 }{ 4 \epsilon_d \left( \omega_{\infty,m} ^{\mathrm{AHE}} \right) ^2 } \right] \right\} \frac{ R ^2 }{ m ^2 } + \omega_p \, \mathcal{O} \left( \frac{ 1 }{ m ^3 } \right) .
\end{multline}
The asymptotes coincide with those of the AHE alone, split according to the relative sign of $ m $ and $ G_z $, while the product $ G_z W $ contributes a term linear in $ q_z $, which makes the two directions of propagation along the axis inequivalent also in this regime.

\begin{figure}[htbp]
    \centering
    \includegraphics[width=0.9\textwidth]{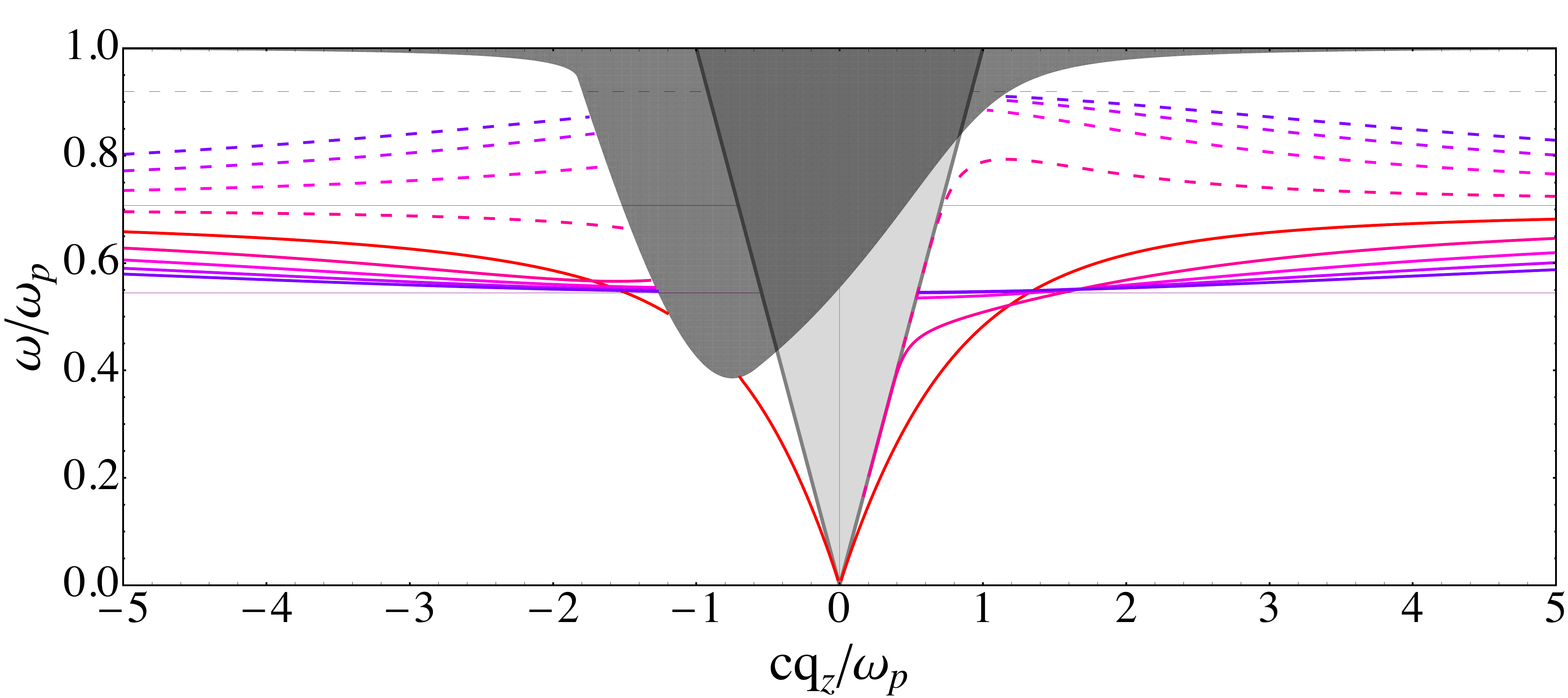}
    \caption{%
        \justifying Numerical SPP dispersion relations at the cylindrical interface with both effects, for $ \omega_p R / c = 1 $, $ c \, G_z / \omega_p = 0.75 $ and $ c \, W / \omega_p = 1.5 $, for both signs of $ q_z $. Solid curves: $ m \leq 0 $; dashed curves: $ m > 0 $. The $ m = 0 $ mode is drawn in red, while the modes with $ \left| m \right| = 1 , 2 , 3 , 4 $ are drawn in magenta, fuchsia, violet and dark violet, respectively. Thin horizontal red line: $ \omega_\infty $, Eq.~\eqref{eq:omegainfty}. Thin horizontal dark violet lines: large-OAM asymptotes $ \omega_{\infty,m} ^{\mathrm{AHE}} $, Eq.~\eqref{eq:asymAHEm}, solid for $ m < 0 $ and dashed for $ m > 0 $. Light grey: dielectric light cone, Eq.~\eqref{eq:lightcone}; dark grey: forbidden region, Eq.~\eqref{eq:forbiddenregion}.
    }
    \label{fig:figure8}
\end{figure}

Figure~\ref{fig:figure8} shows the resulting dispersion relations. The CME modifies the crossings found for the AHE alone: for $ q_z < 0 $, the modes with $ m < 0 $ no longer cross each other, and a single intersection remains, with the $ m = 0 $ mode, whereas for $ q_z > 0 $ several crossings are present.

The symmetries of the dispersion relation are
\begin{equation}
    \omega_m ^{\mathrm{AHECME}} \left( q_z, G_z , W \right) = \omega_m ^{\mathrm{AHECME}} \left( - q_z, G_z , - W \right) = \omega_{-m} ^{\mathrm{AHECME}} \left( - q_z, - G_z , W \right) ,
\end{equation}
which combine those of the two effects taken separately. Unlike the CME alone, the simultaneous change of sign of $ q_z $ and $ m $ is no longer a symmetry, unless $ G_z $ changes sign too: the four modes $ \left( \pm q_z , \pm m \right) $ are therefore in general nondegenerate.

\section{Discussion and Conclusions}
\label{sec:Conclusions}

We have studied SPPs at the interface between a conductor governed by axion electrodynamics and an ordinary dielectric, describing the anomalous response of the conductor through two phenomenological parameters: the axial vector $ \mathbf{G} $, responsible for the AHE, and the pseudoscalar $ W $, accounting for the CME. We have considered the planar geometry in the Voigt, Faraday and perpendicular configurations, and the cylindrical geometry with $ \mathbf{G} $ along the axis. In both, we have derived in closed form the exact eigenmodes of the conductor and, from them, the exact secular equations, which we have solved numerically, identifying the symmetries of the dispersion relations. We have then obtained asymptotic expansions of the frequencies, which make these symmetries explicit, show how the two parameters enter, and capture the propagation features of the modes.

A first question is whether the surface modes have real frequencies at all. In the planar geometry, real solutions are protected by the antiunitary symmetries $ \mathfrak{T} \, \mathfrak{R}_{y,\pi} $ and $ \mathfrak{T} \, \mathfrak{M}_z $, which leave $ q_z $ unchanged and map a solution $ \omega $ into $ \omega ^\star $. As summarised in Table~\ref{tab:antiunitary}, at least one of them survives in the Voigt and Faraday configurations, whereas in the perpendicular configuration the AHE breaks the first and the CME the second, and the frequencies become complex.

\begin{table}[htbp]
\centering
\renewcommand{\arraystretch}{1.4}
\setlength{\tabcolsep}{10pt}
\begin{tabular}{cccc}
    \hline
    Configuration & $ \mathfrak{T} \, \mathfrak{R}_{y,\pi} $ & $ \mathfrak{T} \, \mathfrak{M}_z $ & Real $ \omega $ \\
    \hline
    Voigt CME ($ \mathbf{G} \parallel \mathbf{e}_x $, $ W $) & Preserved & Broken & Yes \\
    Faraday CME ($ \mathbf{G} \parallel \mathbf{e}_z $, $ W $) & Preserved & Broken & Yes \\
    Perpendicular CME ($ \mathbf{G} \parallel \mathbf{e}_y $, $ W $) & Broken if $ \mathbf{G} \neq \mathbf{0} $ & Broken if $ W \neq 0 $ & No \\
    \hline
\end{tabular}
\caption{\justifying Antiunitary symmetries of the planar configurations with the CME, and reality of the SPP frequencies. With the AHE alone, at least one of the two symmetries is preserved in all three configurations.}
\label{tab:antiunitary}
\end{table}

Since any normal component of $ \mathbf{G} $ breaks $ \mathfrak{T} \, \mathfrak{R}_{y,\pi} $, real solutions are no longer protected when $ W \neq 0 $, even for a small tilt out of the plane of the interface, and the freely propagating surface modes are expected to be suppressed in this case as well.

Whenever the frequencies are real, the propagation along the interface is characterised by the reciprocity of the dispersion relation and by the transformations of the parameters that leave it invariant. In the planar geometry, these symmetries are summarised in Table~\ref{tab:planar}.

\begin{table}[htbp]
\centering
\renewcommand{\arraystretch}{1.4}
\setlength{\tabcolsep}{10pt}
    \begin{tabular}{ccr@{}l}
        \hline
        Configuration & Reciprocity & \multicolumn{2}{c}{Symmetry relations} \\
        \hline
        Voigt & Broken & $ \left( q_z, \mathbf{G} \right) \to {} $ & $ \left( - q_z, - \mathbf{G} \right) $ \\
        Faraday & Preserved & $ \mathbf{G} \to {} $ & $ - \mathbf{G} $ \\
        Perpendicular & Preserved & $ \mathbf{G} \to {} $ & $ - \mathbf{G} $ \\
        CME & Preserved & $ W \to {} $ & $ - W $ \\
        \hline
        \multirow{2}{*}{Voigt CME} & \multirow{2}{*}{Broken} & \multirow{2}{*}{$ \left( q_z, \mathbf{G}, W \right) \to {} $} & $ \left( - q_z, - \mathbf{G}, W \right) $ \\
        & & & $ \left( q_z, \mathbf{G}, - W \right) $ \\
        \hline
        \multirow{2}{*}{Faraday CME} & \multirow{2}{*}{Broken} & \multirow{2}{*}{$ \left( q_z, \mathbf{G}, W \right) \to {} $} & $ \left( - q_z, - \mathbf{G}, W \right) $ \\
        & & & $ \left( - q_z, \mathbf{G}, - W \right) $ \\
        \hline
    \end{tabular}
\caption{\justifying Reciprocity and symmetry relations of the planar dispersion relations. The perpendicular configuration with the CME is not included, since it admits no freely propagating SPPs.}
\label{tab:planar}
\end{table}

The cases with the AHE alone, included in the tables for completeness, are known from Refs.~\cite{Hofmann2016,OskouiAbdol2020}; the new results concern the role of $ W $. The CME alone leaves the dispersion relation reciprocal and enters only through $ W ^2 $, but it reverses the group velocity at large wavevector for $ | W | > 2 \sqrt{ \epsilon_d } \, \omega_\infty / c $, and sets the upper bound $ | W | < 2 \sqrt{ \epsilon_\infty } \, \omega_p / c $ for the existence of SPPs. In the Voigt configuration, $ W $ makes the threshold for this reversal different for the two directions of propagation, without adding a nonreciprocity of its own; for a strong AHE, only one direction can be reversed. In the Faraday configuration, instead, the coexistence of $ \mathbf{G} $ and $ W $ breaks reciprocity, which neither of them breaks alone, through a term proportional to $ G_{\mathrm{F}} W / q_z $. The resulting dispersion relations need no longer be monotonic, and at large wavevector the group velocity of both branches has the opposite sign to that of $ G_{\mathrm{F}} W $.

In the cylindrical geometry, the OAM provides a second label for the modes, and reciprocity must be considered separately in $ q_z $ and in $ m $, as summarised in Table~\ref{tab:cylindrical}. While the AHE alone separates the modes with opposite OAM~\cite{Peluso2025WG}, the CME alone, which is reciprocal at the planar interface, couples the linear and orbital momenta of the modes, breaking reciprocity in $ q_z $ and in $ m $ separately, but preserving their simultaneous reversal. When both effects are present, no reversal of $ q_z $, of $ m $ or of both is a symmetry: the four modes $ \left( \pm q_z , \pm m \right) $ are in general nondegenerate, and even the fundamental mode $ m = 0 $, reciprocal for either effect alone, becomes nonreciprocal.

\begin{table}[htbp]
\centering
\renewcommand{\arraystretch}{1.4}
\setlength{\tabcolsep}{10pt}
    \begin{tabular}{cccr@{}l}
        \hline
        Configuration & Reciprocity in $ q_z $ & Reciprocity in $ m $ & \multicolumn{2}{c}{Symmetry relations} \\
        \hline
        AHE & Preserved & Broken & $ \left( m, \mathbf{G} \right) \to {} $ & $ \left( - m, - \mathbf{G} \right) $ \\
        \multirow{2}{*}{CME} & \multirow{2}{*}{Broken} & \multirow{2}{*}{Broken} & \multirow{2}{*}{$ \left( q_z, m, W \right) \to {} $} & $ \left( q_z, - m, - W \right) $ \\
        & & & & $ \left( - q_z, m, - W \right) $ \\
        \hline
        \multirow{2}{*}{AHE and CME} & \multirow{2}{*}{Broken} & \multirow{2}{*}{Broken} & \multirow{2}{*}{$ \left( q_z, m, \mathbf{G}, W \right) \to {} $} & $ \left( - q_z, m, \mathbf{G}, - W \right) $ \\
        & & & & $ \left( - q_z, - m, - \mathbf{G}, W \right) $ \\
        \hline
    \end{tabular}
\caption{\justifying Reciprocity in $ q_z $ and in $ m $, and symmetry relations of the dispersion relations in the cylindrical geometry, with $ \mathbf{G} $ along the axis.}
\label{tab:cylindrical}
\end{table}

Behind these results lies a simple symmetry consideration. The sign of $ W $ affects the dispersion relations only through a combination of the same symmetry, built from $ \mathbf{G} $ and $ \mathbf{q} $ in both the Faraday configuration and the cylindrical geometry, and from $ m $ and $ q_z $ in the cylinder alone; otherwise, as in the Voigt configuration and for the CME alone at the planar interface, they are even in $ W $. Through $ \mathbf{G} $, a large and known AHE amplifies a small $ W $ into a nonreciprocity that is largest at moderate $ | q_z | $, whereas the coupling to the OAM requires no AHE.

The application of the theory to real systems requires including losses, which broaden the dispersion relations, nonlocal effects at large wavevector and possible contributions localised at the interface, as well as identifying a mechanism that sustains the CME out of equilibrium. These ingredients affect the magnitude of the effects, but not the symmetry relations in Tables~\ref{tab:planar} and~\ref{tab:cylindrical}, which follow from the transformation properties of $ \mathbf{G} $ and $ W $ alone. Whether the splittings induced by $ W $ can be resolved in a given medium, for instance in magnetic Weyl semimetals with a large anomalous Hall response~\cite{Burkov2014,Kotov2018}, requires a quantitative estimate beyond the scope of this work. Natural extensions include the role of $ W $ in thin films~\cite{Kotov2018,Tamaya2019} and in conical tips, where the OAM-selective nanofocusing produced by $ \mathbf{G} $~\cite{Peluso2025NF} could be modified by the coupling between linear and orbital momenta found here. In microscopic models of the dynamical anomalous Hall and chiral magnetic responses~\cite{Chen2013,Ma2015,Zhong2016,Kotov2018}, frequency-dependent $ \mathbf{G} $ and $ W $ would modify the asymptotes and the thresholds, while nonuniform ones, as at the domain walls between regions with different magnetisation, would allow surface modes guided along the walls~\cite{ZyuzinZyuzin2015}.

To conclude, our analysis identifies axion media as a platform in which the chiral magnetic effect, through its interplay with the anomalous Hall effect and with the orbital angular momentum, controls the nonreciprocity of surface plasmon polaritons, with prospects for nonreciprocal plasmonics driven by the intrinsic response of matter.

\ack{I thank A. De Martino for discussions.}

\funding{This work has been supported by the fellowship DM 118/2023 - Inv. 4.1, project ``Light--matter interactions in topological semimetals,'' CUP E14D23001640006, PNRR.}

\data{The numerical data shown in the figures of this paper will be available on Zenodo upon publication.}

\appendix

\section{Boundary Conditions}
\label{app:Boundary}

The boundary conditions at the interface follow from the integration of Maxwell's equations across it, over a thin pillbox for the divergence equations and over a thin loop for the curl equations~\cite{Landau1984}. As the interface carries no surface charges or currents, the matching conditions read
\begin{align}
    D_{\mathrm{c},\perp} - D_{\mathrm{d},\perp} &= 0 , &
    B_{\mathrm{c},\perp} - B_{\mathrm{d},\perp} &= 0 , \\
    \mathbf{E}_{\mathrm{c},\parallel} - \mathbf{E}_{\mathrm{d},\parallel} &= \mathbf{0} , &
    \mathbf{H}_{\mathrm{c},\parallel} - \mathbf{H}_{\mathrm{d},\parallel} &= \mathbf{0} ,
\end{align}
where the subscripts $ \mathrm{c} $ and $ \mathrm{d} $ refer to the conductor and to the dielectric, $ \perp $ and $ \parallel $ denote the components normal and tangential to the interface, and $ \mathbf{H} = \mathbf{B} / \mu_0 $ in both media, which are considered nonmagnetic. In the dielectric $ \mathbf{D}_{\mathrm{d}} = \varepsilon_0 \epsilon_d \mathbf{E}_{\mathrm{d}} $, while in the conductor all the contributions to the charge and current densities, ordinary and anomalous, are collected in the displacement field
\begin{equation}
    \mathbf{D}_{\mathrm{c}} = \varepsilon_0 \epsilon \left( \omega \right) \mathbf{E}_{\mathrm{c}} + \frac{ \varepsilon_0 c }{ \mathrm{i} \omega } \mathbf{G} \times \mathbf{E}_{\mathrm{c}} + \frac{ \varepsilon_0 c ^2 W }{ \omega ^2 } \nabla \times \mathbf{E}_{\mathrm{c}} .
\end{equation}
With this definition, Gauss's law and the Amp\`ere--Maxwell law take the standard forms $ \nabla \cdot \mathbf{D}_{\mathrm{c}} = 0 $ and $ \nabla \times \mathbf{H}_{\mathrm{c}} = - \mathrm{i} \omega \mathbf{D}_{\mathrm{c}} $. For monochromatic fields, the conditions on the normal components follow from those on the tangential ones through Faraday's law and the Amp\`ere--Maxwell law, so that only the matching of the tangential components~\eqref{eq:boundary} needs to be imposed.

\section{Technical Details on Planar Interface}
\label{app:Planar}

In the planar geometry the fields depend on $ z $ through $ \mathit{e} ^{ \mathrm{i} q_z z } $, are uniform along $ x $ and decay away from the interface $ y = 0 $. In each medium we first determine the evanescent eigenmodes, then impose the boundary conditions and derive the secular equation, whose asymptotic forms give the expansions of Section~\ref{sec:Planar}.

\subsection{Dielectric and conventional conductor}
\label{app:PlanarDiel}

In an isotropic medium the transversality condition reduces to $ \mathbf{q} \cdot \mathbf{E} = 0 $, and the field of a plane wave can be decomposed into a TE and a TM component with respect to the $ z $ axis,
\begin{equation}
    \widetilde{\mathbf{E}} =
    C_{\mathrm{TE}}
    \begin{bmatrix}
        q_y \\[8pt]
        - q_x \\[8pt]
        0
    \end{bmatrix}
    +
    C_{\mathrm{TM}}
    \begin{bmatrix}
        q_x \\[8pt]
        q_y \\[8pt]
        - \dfrac{ q_x ^2 + q_y ^2 }{ q_z }
    \end{bmatrix} ,
\end{equation}
where $ C_{\mathrm{TE}} $ and $ C_{\mathrm{TM}} $ are arbitrary amplitudes. In the dielectric, for $ \mathbf{q} = \left( 0 , \mathrm{i} \kappa_d , q_z \right) $, the dispersion relation $ \mathbf{q} \cdot \mathbf{q} = k_0 ^2 \epsilon_d $ gives the inverse decay length
\begin{equation}
    \kappa_d = \sqrt{ q_z ^2 - k_0 ^2 \epsilon_d } ,
\end{equation}
with $ k_0 = \omega / c $. Rescaling the amplitudes, the electric field and the magnetic field obtained from Faraday's law read
\begin{align}
    \widetilde{\mathbf{E}}_{\mathrm{TEd}} + \widetilde{\mathbf{E}}_{\mathrm{TMd}} &=
    \frac{ \mathrm{i} \omega \mathcal{A}_{\mathrm{TEd}} }{ q_z }
    \begin{bmatrix}
        \mathrm{i} \kappa_d \\[8pt]
        0 \\[8pt]
        0
    \end{bmatrix}
    + \frac{ \mathrm{i} \omega \mathcal{A}_{\mathrm{TMd}} }{ q_z }
    \begin{bmatrix}
        0 \\[8pt]
        - \kappa_d \\[8pt]
        \dfrac{ \mathrm{i} \kappa_d ^2 }{ q_z }
    \end{bmatrix}, &
    \widetilde{\mathbf{B}}_{\mathrm{TEd}} + \widetilde{\mathbf{B}}_{\mathrm{TMd}} &=
    \mathcal{A}_{\mathrm{TEd}}
    \begin{bmatrix}
        0 \\[8pt]
        - \kappa_d \\[8pt]
        \dfrac{ \mathrm{i} \kappa_d ^2 }{ q_z }
    \end{bmatrix}
    + \frac{ \omega ^2 \epsilon_d }{ c ^2 q_z ^2 } \mathcal{A}_{\mathrm{TMd}}
    \begin{bmatrix}
        \mathrm{i} \kappa_d \\[8pt]
        0 \\[8pt]
        0
    \end{bmatrix} .
\end{align}

The expressions for the vectors of a conventional conductor are obtained from the previous ones by the simple substitutions $ \epsilon_d \rightarrow \epsilon \left( \omega \right) $ and $ \kappa_d \rightarrow - \kappa_\epsilon $, where
\begin{equation}
    \kappa_\epsilon = \sqrt{ q_z ^2 - k_0 ^2 \epsilon \left( \omega \right) }
\end{equation}
is the inverse decay length in the conductor. Imposing the continuity of $ E_x $, $ E_z $, $ H_x $ and $ H_z $ at $ y = 0 $, the TE and TM components decouple. The TE system admits only the trivial solution, while the TM one has a nontrivial solution if
\begin{equation}
    \frac{ \kappa_\epsilon }{ \epsilon \left( \omega \right) } + \frac{ \kappa_d }{ \epsilon_d } = 0,
\end{equation}
which requires $ \epsilon \left( \omega \right) < 0 $. With the Drude permittivity~\eqref{eq:Drude}, it gives the dispersion relation of the conventional SPP,
\begin{equation}
    \omega_{\mathrm{SPP}} \left( q_z \right) = \sqrt{ \frac{ 1 }{ 2 } \left[ \left( \frac{ \epsilon_d + \epsilon_\infty }{ \epsilon_d \, \epsilon_\infty } \right) c ^2 q_z ^2 + \omega_p ^2 \right] - \sqrt{ \frac{ 1 }{ 4 } \left[ \left( \frac{ \epsilon_d + \epsilon_\infty }{ \epsilon_d \, \epsilon_\infty } \right) c ^2 q_z ^2 + \omega_p ^2 \right] ^2 - \frac{ c ^2 q_z ^2 \omega_p ^2 }{ \epsilon_d } } } ,
\end{equation}
which starts from the light line of the dielectric,
\begin{equation}
    \omega_{\mathrm{SPP}} \left( q_z \to 0 \right) = \frac{ c | q_z | }{ \sqrt{ \epsilon_d } } - \frac{ c ^3 | q_z | ^3 }{ 2 \sqrt{ \epsilon_d } \epsilon_\infty \omega_p ^2 } + \omega_p \, \mathcal{O} \left( \frac{ c ^5 q_z ^5 }{ \omega_p ^5 } \right) .
\end{equation}
At large wavevector $ \kappa_\epsilon \simeq \kappa_d \simeq | q_z | $, and the secular equation reduces at leading order to
\begin{equation}
    \epsilon \left( \omega \right) + \epsilon_d = 0 ,
\end{equation}
whose solution is the surface plasmon frequency $ \omega_\infty $, Eq.~\eqref{eq:omegainfty}, approached from below as
\begin{equation}
    \omega_{\mathrm{SPP}} \left( q_z \to \pm \infty \right) = \omega_\infty - \frac{ \epsilon_d ^2 \, \omega_\infty ^3 }{ 2 c ^2 \left( \epsilon_\infty + \epsilon_d \right) } \frac{ 1 }{ q_z ^2 } + \omega_p \, \mathcal{O} \left( \frac{ \omega_p ^4 }{ c ^4 q_z ^4 } \right) .
\end{equation}

\subsection{Anomalous conductor}
\label{app:PlanarCond}

In the conductor we introduce the right-handed orthonormal basis $ \left( \mathbf{e}_{\mathrm{T1}} , \mathbf{e}_{\mathrm{T2}} , \mathbf{e}_G \right) $, with $ \mathbf{G} = G \, \mathbf{e}_G $, and decompose the wavevector into its components $ q_{\mathrm{T1}} $, $ q_{\mathrm{T2}} $ and $ q_G $, with $ q ^2 = \mathbf{q} \cdot \mathbf{q} $. In this basis, the wave equation~\eqref{eom} for a plane wave reads
\begin{equation}
    \begin{bmatrix}
        q ^2 - k_0 ^2 \epsilon - q_\mathrm{T1} ^2 &
        - q_\mathrm{T1} q_\mathrm{T2} - \mathrm{i} k_0 G + \mathrm{i} W q_G &
        - q_\mathrm{T1} q_G - \mathrm{i} W q_\mathrm{T2} \\[8pt]
        - q_\mathrm{T1} q_\mathrm{T2} + \mathrm{i} k_0 G - \mathrm{i} W q_G &
        q ^2 - k_0 ^2 \epsilon - q_\mathrm{T2} ^2 &
        - q_\mathrm{T2} q_G + \mathrm{i} W q_\mathrm{T1} \\[8pt]
        - q_\mathrm{T1} q_G + \mathrm{i} W q_\mathrm{T2} &
        - q_\mathrm{T2} q_G - \mathrm{i} W q_\mathrm{T1} &
        q ^2 - k_0 ^2 \epsilon - q_G ^2
    \end{bmatrix}
    \begin{bmatrix}
        E_\mathrm{T1} \\[8pt]
        E_\mathrm{T2} \\[8pt]
        E_G
    \end{bmatrix}
    = 0.
\end{equation}

Its determinant vanishes on the bulk dispersion relation~\eqref{eq:bulk_imp_dispersion_relation}. On each solution, the field orthogonal to $ \mathbf{G} $ can be parametrised as
\begin{align}
    E_\mathrm{T1} &\propto q_\mathrm{T2} + \mathrm{i} \Gamma q_\mathrm{T1} , &
    E_\mathrm{T2} &\propto - q_\mathrm{T1} + \mathrm{i} \Gamma q_\mathrm{T2} ,
\end{align}
that is, as the superposition of a part orthogonal to both $ \mathbf{G} $ and $ \mathbf{q} $ and one in the plane they span, coupled through the hybridisation parameter $ \Gamma $. The component along $ \mathbf{G} $ follows from the corresponding row of the wave equation~\eqref{eom}, and $ \Gamma $ from the transversality condition~\eqref{divE},
\begin{align}
    E_G &\propto \mathrm{i} \Gamma \left( q_G - W / \Gamma \right) \frac{ q_\perp ^2 }{ q_\perp ^2 - k_0 ^2 \epsilon } , \\[8pt]
    \Gamma &= \frac{ G \left( q_\perp ^2 - k_0 ^2 \epsilon \right) + W q_G k_0 \epsilon }{ k_0 \epsilon \left( q ^2 - k_0 ^2 \epsilon \right) } ,
\end{align}
where $ q_\perp ^2 = q_{\mathrm{T1}} ^2 + q_{\mathrm{T2}} ^2 $. For an evanescent wave, the hybridisation parameter takes one value for each inverse decay length, $ \Gamma_\pm $ for $ \kappa_\pm $, and depends on both $ \mathbf{G} $ and $ W $. In the following we specialise these expressions to the three configurations.

\subsubsection{Faraday CME}
\label{app:PlanarFaraday}

For $ \mathbf{G} = G_{\mathrm{F}} \mathbf{e}_z $ and $ \mathbf{q} = \left( 0 , - \mathrm{i} \kappa_\pm , q_z \right) $, we have $ q_G = q_z $ and $ q_\perp ^2 = - \kappa_\pm ^2 $, and the bulk dispersion relation becomes
\begin{equation}
    \left( q_z ^2 - \kappa_\pm ^2 - k_0 ^2 \epsilon \right) ^2 \epsilon - G_{\mathrm{F}} ^2 \left( k_0 ^2 \epsilon + \kappa_\pm ^2 \right) - W ^2 \left( q_z ^2 - \kappa_\pm ^2 \right) \epsilon + 2 k_0 \epsilon W G_{\mathrm{F}} q_z  = 0 ,
\end{equation}
a quadratic equation for $ \kappa_\pm ^2 $, whose roots with positive real part are
\begin{equation}
    \kappa_\pm = \sqrt{ q_z ^2 - k_0 ^2 \epsilon - \frac{ W ^2 }{ 2 } + \frac{ G_{\mathrm{F}} ^2 }{ 2 \epsilon } \pm \sqrt{ \Delta } } ,
\end{equation}
where
\begin{equation}
    \Delta = \left( G_{\mathrm{F}} ^2 + 4 q_z ^2 \epsilon \right) \frac{ G_{\mathrm{F}} ^2 }{ 4 \epsilon ^2 } + \left( W k_0 ^2 \epsilon + \frac{ W ^3 }{ 4 } - 2 G_{\mathrm{F}} q_z k_0 - \frac{ G_{\mathrm{F}} ^2 W }{ 2 \epsilon } \right) W .
\end{equation}

The term $ - 2 G_{\mathrm{F}} q_z k_0 W $ in $ \Delta $ is the only one odd in $ q_z $. The corresponding hybridisation parameters are
\begin{equation}
    \Gamma_\pm = \frac{ G_{\mathrm{F}} ^2 q_z + W \epsilon \left( W q_z - 2 G_{\mathrm{F}} k_0 \right) \pm 2 q_z \epsilon \sqrt{ \Delta }
    }{ 2 k_0 \epsilon \left( G_{\mathrm{F}} q_z - W k_0 \epsilon \right) } ,
\end{equation}
and the fields of the two eigenmodes in the conductor are
\begin{align}
    \widetilde{\mathbf{E}}_\pm
    & =
    \frac{ \mathrm{i} \omega \mathcal{A}_\pm }{ q_z }
    \left\{
    \begin{bmatrix}
        - \mathrm{i} \kappa_\pm \\[16pt]
        0 \\[16pt]
        0
    \end{bmatrix}
    +
    \mathrm{i} \Gamma_\pm
    \begin{bmatrix}
        0 \\[16pt]
        - \mathrm{i} \kappa_\pm \\[16pt]
        \dfrac{ \left( q_z - W / \Gamma_\pm \right) \kappa_\pm ^2 }{ \left( k_0 ^2 \epsilon + \kappa_\pm ^2  \right) }
    \end{bmatrix}
    \right\},
    &
    \widetilde{\mathbf{B}}_\pm
    & =
    \mathcal{A}_\pm
    \left\{
    \begin{bmatrix}
        0 \\[16pt]
        \kappa_\pm \\[16pt]
        \dfrac{ \mathrm{i} \kappa_\pm ^2 }{ q_z }
    \end{bmatrix}
    +
    \frac{ \Gamma_\pm k_0 ^2 \epsilon + \kappa_\pm ^2 W / q_z }{ \left( k_0 ^2 \epsilon + \kappa_\pm ^2 \right) }
    \begin{bmatrix}
        - \mathrm{i} \kappa_\pm \\[16pt]
        0 \\[16pt]
        0
    \end{bmatrix}
    \right\} ,
\end{align}
where the first term is TE-like and the second TM-like. Imposing the continuity of $ E_x $, $ E_z $, $ H_x $ and $ H_z $ at $ y = 0 $ and eliminating the amplitudes $ \mathcal{A}_\pm $, $ \mathcal{A}_{\mathrm{TEd}} $ and $ \mathcal{A}_{\mathrm{TMd}} $, we obtain the secular equation
\begin{equation}
    \sum_{ \sigma = \pm } \sigma \left( 1 + \dfrac{ \kappa_{ -\sigma } }{ \kappa_d } \right) \left[ \left( \dfrac{ \Gamma_{ \sigma } q_z ^2 - W q_z }{ k_0 ^2 \epsilon + \kappa_{ \sigma } ^2 } \right) \dfrac{ \kappa_{ \sigma } }{ \kappa_d } k_0 ^2 \epsilon_d + \left( \dfrac{ \Gamma_{ \sigma } q_z ^2 k_0 ^2 \epsilon + q_z \kappa_{\sigma} ^2 W }{ k_0 ^2 \epsilon + \kappa_{ \sigma } ^2 } \right) \right] = 0.
\end{equation}

At large wavevector, expanding $ \kappa_\pm $, $ \kappa_d $ and $ \Gamma_\pm $ in powers of $ 1 / q_z $ (here and in the following), the secular equation becomes equivalent to
\begin{equation}
    \epsilon \left( \omega \right) + \epsilon_d = \frac{ G_{\mathrm{F}} W }{ 4 k_0 } \frac{ 1 }{ q_z } - \left( k_0 ^2 \epsilon_d ^2 + \frac{ G_{\mathrm{F}} ^2 }{ 4 } - \frac{ \epsilon_d W ^2 }{ 4 } \right) \frac{ 1 }{ q_z ^2 } + \mathcal{O} \left( \frac{ \omega_p ^3 }{ c ^3 q_z ^3 } \right) ,
\end{equation}
which is of the form~\eqref{eq:asymptotic_form} and, inverted as in Appendix~\ref{app:MathAsym}, gives the expansion of $ \omega_{\mathrm{FCME}} $ in Section~\ref{sec:PlanarFaraday}.

Near the light line of the dielectric, instead, we follow the conventional SPP and look for solutions with $ \omega \to c | q_z | / \sqrt{ \epsilon_d } $ as $ q_z \to 0 $, where $ \kappa_d \to 0 $ and $ | \epsilon \left( \omega \right) | \to \infty $. Solved order by order for $ q_z ^2 $, the secular equation becomes equivalent to
\begin{equation}
    q_z ^2 = k_0 ^2 \epsilon_d - \frac{ \epsilon_d ^2 }{ \epsilon ^2 } \left( k_0 ^2 \epsilon + \frac{ W ^2 }{ 4 } \right) + \mathcal{O} \left( q_z ^6 \right) .
\end{equation}

All the terms involving $ \mathbf{G} $ in $ \kappa_\pm $ and $ \Gamma_\pm $ are suppressed by inverse powers of $ \epsilon $ or by powers of $ k_0 $ and $ q_z $, and therefore do not contribute at this order, as also happens in the Voigt and perpendicular configurations. Inverted at small $ q_z $ with $ \epsilon \simeq - \epsilon_\infty \omega_p ^2 / \omega ^2 $, this relation gives the expansions near the light line of Sections~\ref{sec:PlanarAHE} and~\ref{sec:PlanarCME}, whose cubic correction is proportional to $ 1 - c ^2 W ^2 / \left( 4 \epsilon_\infty \omega_p ^2 \right) $. The combination $ k_0 ^2 \epsilon + W ^2 / 4 $ originates from the inverse decay lengths $ \kappa_\pm $ of the conductor, so that the resulting bound on $ W $ does not involve $ \epsilon_d $. The cases in which only one of the two effects is present follow as particular cases.

\paragraph{AHE alone}

For $ W = 0 $ the bulk dispersion relation reduces to
\begin{equation}
    \left( q_z ^2 - \kappa_\pm ^2 - k_0 ^2 \epsilon \right) ^2 \epsilon - G_{\mathrm{F}} ^2 \left( k_0 ^2 \epsilon + \kappa_\pm ^2 \right) = 0 ,
\end{equation}
whose roots are
\begin{equation}
    \kappa_\pm = \sqrt{ q_z ^2 - k_0 ^2 \epsilon + \frac{ G_{\mathrm{F}} ^2 }{ 2 \epsilon } \pm \sqrt{ \left( G_{\mathrm{F}} ^2 + 4 q_z ^2 \epsilon \right) \frac{ G_{\mathrm{F}} ^2 }{ 4 \epsilon ^2 } } }.
\end{equation}

The hybridisation parameters become
\begin{align}
    \Gamma_\pm &= \frac{ G_{\mathrm{F}} }{ 2 k_0 \epsilon } \pm \frac{ 1 }{ k_0 G_{\mathrm{F}} } \sqrt{ \left( G_{\mathrm{F}} ^2 + 4 q_z ^2 \epsilon \right) \dfrac{ G_{\mathrm{F}} ^2 }{ 4 \epsilon ^2 } }, &
    \Gamma_\pm ^{-1} &= \frac{ k_0 ^2 \epsilon }{ k_0 ^2 \epsilon + \kappa_\pm ^2 } \Gamma_\pm,
\end{align}
where the second relation, used to simplify the fields, follows from the bulk dispersion relation. All the quantities depend on $ q_z $ only through $ q_z ^2 $, which makes the propagation reciprocal. Since $ \Gamma_\pm $ is odd in $ G_{\mathrm{F}} $ and $ \kappa_\pm $ is even, the dispersion relation is also even in $ G_{\mathrm{F}} $.

\paragraph{CME alone}

For $ \mathbf{G} = \mathbf{0} $ the bulk dispersion relation becomes
\begin{equation}
    \left( q_z ^2 - \kappa_\pm ^2 - k_0 ^2 \epsilon \right) ^2 - W ^2 \left( q_z ^2 - \kappa_\pm ^2 \right) = 0 ,
\end{equation}
with roots
\begin{equation}
    \kappa_\pm = \sqrt{ q_z ^2 - k_0 ^2 \epsilon - \frac{ W ^2 }{ 2 } \pm \sqrt{ \left( k_0 ^2 \epsilon + \frac{ W ^2 }{ 4 } \right) W ^2 } } ,
\end{equation}
and hybridisation parameters
\begin{align}
    \Gamma_\pm &= - \frac{ q_z }{ k_0 ^2 \epsilon } \left[ \frac{ W }{ 2 } \pm \frac{ 1 }{ W } \sqrt{ \left( k_0 ^2 \epsilon + \frac{ W ^2 }{ 4 } \right) W ^2 } \right], &
    \Gamma_\pm ^{-1} &= \frac{ q_z ^2 - \kappa_\pm ^2 }{ q_z ^2 } \Gamma_\pm .
\end{align}

Since the conductor is isotropic, these expressions do not depend on the configuration, and describe the CME alone of Section~\ref{sec:PlanarCME}. The corresponding secular equation is that of the Faraday configuration with $ G_{\mathrm{F}} = 0 $. As $ \kappa_\pm $ and $ \kappa_d $ depend on $ q_z $ and $ W $ only through $ q_z ^2 $ and $ W ^2 $, while $ \Gamma_\pm $ is odd in both, it only changes sign under $ q_z \to - q_z $ or $ W \to - W $: the dispersion relation is reciprocal and even in $ W $.

\subsubsection{Voigt CME}
\label{app:PlanarVoigt}

For $ \mathbf{G} = G_{\mathrm{V}} \mathbf{e}_x $ the wavevector has no component along $ \mathbf{G} $, $ q_G = 0 $, and the term $ W \left( \mathbf{q} \cdot \mathbf{G} \right) $ vanishes. The bulk dispersion relation is then even in $ q_z $, and its roots are
\begin{equation}
    \kappa_\pm = \sqrt{ q_z ^2 - k_0 ^2 \epsilon - \frac{ W ^2 }{ 2 } + \frac{ G_{\mathrm{V}} ^2 }{ 2 \epsilon } \pm \sqrt{ \Delta } },
\end{equation}
where
\begin{equation}
    \Delta = \frac{ G_{\mathrm{V}} ^4 }{ 4 \epsilon ^2 } + \left( W k_0 ^2 \epsilon + \frac{ W ^3 }{ 4 } - \frac{ G_{\mathrm{V}} ^2 W }{ 2 \epsilon } \right) W .
\end{equation}

In this configuration the hybridisation parameter is the same for both eigenmodes,
\begin{equation}
    \Gamma = \frac{ G_{\mathrm{V}} }{ k_0 \epsilon } ,
\end{equation}
and the fields in the conductor are
\begin{align}
    \widetilde{\mathbf{E}}_\pm
    & =
    \frac{ \mathrm{i} \omega \mathcal{A}_\pm }{ q_z }
    \left\{
    \begin{bmatrix}
        0 \\[16pt]
        q_z \\[16pt]
        \mathrm{i} \kappa_\pm
    \end{bmatrix}
    +
    \mathrm{i}
    \begin{bmatrix}
        \dfrac{ W \left( q_z ^2 - \kappa_\pm ^2  \right) }{ k_0 ^2 \epsilon + \kappa_\pm ^2 - q_z ^2 } \\[16pt]
        - \dfrac{ \mathrm{i} G_{\mathrm{V}} }{ k_0 \epsilon } \kappa_\pm \\[16pt]
        \dfrac{ G_{\mathrm{V}} }{ k_0 \epsilon } q_z
    \end{bmatrix}
    \right\},
    &
    \widetilde{\mathbf{B}}_\pm
    & =
    - \mathcal{A}_\pm
    \left\{
    \begin{bmatrix}
        \dfrac{ \mathrm{i} \left( q_z ^2 - \kappa_\pm ^2  \right) }{ q_z } \\[16pt]
        0 \\[16pt]
        0
    \end{bmatrix}
    +
    \dfrac{ W \left( q_z ^2 - \kappa_\pm ^2  \right) }{ k_0 ^2 \epsilon + \kappa_\pm ^2 - q_z ^2 }
    \begin{bmatrix}
        0 \\[16pt]
        1 \\[16pt]
        \dfrac{ \mathrm{i} \kappa_\pm }{ q_z }
    \end{bmatrix}
    \right\},
\end{align}
in which the CME mixes the TM wave produced by the AHE with a TE component, of first order in $ W $. The boundary conditions give the secular equation
\begin{equation}
    \sum_{ \sigma = \pm } \sigma \left( \dfrac{ \kappa_d + \kappa_{ \sigma } }{ k_0 ^2 \epsilon + \kappa_{ \sigma } ^2 - q_z ^2 } \right) \left[ \left( \dfrac{ \kappa_{ - \sigma } + \dfrac{ G_{\mathrm{V}} }{ k_0 \epsilon } q_z }{ q_z ^2 - \kappa_{ - \sigma } ^2  } \right) \dfrac{ k_0 ^2 \epsilon_d }{ \kappa_d } + 1 \right] = 0 ,
\end{equation}
which, at large wavevector, takes the form
\begin{equation}
    \left[ \epsilon \left( \omega \right) + \epsilon_d
    \right] - \frac{ \operatorname{sgn} \left( q_z \right) \, G_{\mathrm{V}} } { k_0 }
    = - \frac{ \epsilon_d }{ 4 q_z ^2 }
    \left( 4 k_0 ^2 \epsilon_d - W ^2 \right)
    + \mathcal{O} \left( \frac{ \omega_p ^4 }{ c ^4 q_z ^4 } \right) .
\end{equation}

At leading order, the left-hand side vanishes at the asymptotic frequency $ \omega_{\infty,s} ^{\mathrm{V}} $ of Section~\ref{sec:PlanarAHE}, which depends on $ \operatorname{sgn} \left( q_z \right) $, while $ W $ enters only the correction of order $ 1 / q_z ^2 $.

\paragraph{AHE alone}

For $ W = 0 $ the bulk dispersion relation factorises,
\begin{equation}
    \left( q_z ^2 - k_0 ^2 \epsilon + \frac{ G_{\mathrm{V}} ^2 }{ \epsilon } - \kappa_{\mathrm{V}} ^2 \right) \left( q_z ^2 - k_0 ^2 \epsilon - \kappa_\epsilon ^2 \right) = 0.
\end{equation}

The second factor corresponds to a TE wave, which is not affected by $ G_{\mathrm{V}} $ and does not take part in the surface mode, while the first gives the TM-like wave

\begin{align}
    \widetilde{\mathbf{E}}_{\mathrm{V}} &= \frac{ \mathrm{i} \omega \mathcal{A}_{\mathrm{V}} }{ q_z } \left\{
        \begin{bmatrix}
            0 \\[8pt]
            q_z \\[8pt]
            \mathrm{i} \kappa_{\mathrm{V}}
        \end{bmatrix}
        + \frac{ \mathrm{i} G_{\mathrm{V}} }{ k_0 \epsilon }
        \begin{bmatrix}
            0 \\[8pt]
            - \mathrm{i} \kappa_{\mathrm{V}} \\[8pt]
            q_z
        \end{bmatrix}
    \right\}, &
    \kappa_{\mathrm{V}} &= \sqrt{ q_z ^2 - k_0 ^2 \epsilon + \frac{ G_{\mathrm{V}} ^2 }{ \epsilon } }.
\end{align}

The boundary conditions then reduce to
\begin{equation}
    \frac{ \kappa_{\mathrm{V}} }{ \epsilon } \left( 1 + \frac{ q_z G_{\mathrm{V}} }{ \kappa_{\mathrm{V}} k_0 \epsilon } \right) + \frac{ \kappa_d }{ \epsilon_d } \left( 1 - \frac{ G_{\mathrm{V}} ^2 }{ k_0 ^2 \epsilon ^2 } \right) = 0.
\end{equation}

The term linear in $ q_z G_{\mathrm{V}} $ is the origin of the nonreciprocity of the Voigt configuration.

\subsubsection{Perpendicular CME}
\label{app:PlanarPerpendicular}

For $ \mathbf{G} = G_{\mathrm{P}} \mathbf{e}_y $ the component of the wavevector along $ \mathbf{G} $ is the normal one, $ q_G = - \mathrm{i} \kappa_\pm $, and the bulk dispersion relation acquires a term odd in $ \kappa_\pm $ with an imaginary coefficient,
\begin{equation}
    \left( q_z ^2 - \kappa_\pm ^2 - k_0 ^2 \epsilon \right) ^2 \epsilon - G_{\mathrm{P}} ^2 \left( k_0 ^2 \epsilon - q_z ^2 \right) - W ^2 \left( q_z ^2 - \kappa_\pm ^2 \right) \epsilon - 2 \mathrm{i} k_0 \epsilon W G_{\mathrm{P}} \kappa_\pm  = 0 .
\end{equation}

This is a quartic equation in $ \kappa_\pm $, which is no longer a quadratic in $ \kappa_\pm ^2 $. Its roots, obtained through the resolvent cubic of the quartic~\cite{NISTDLMF}, are
\begin{equation}
    \kappa_\pm = \sqrt{ q_z ^2 - k_0 ^2 \epsilon - \frac{ W ^2 }{ 2 } + \frac{ \mathfrak{a} }{ 4 } \pm \frac{ k_0 G_{\mathrm{P}} W }{ \sqrt{ \mathfrak{a} } } } \pm \frac{ \mathrm{i} }{ 2 } \sqrt{ \mathfrak{a} } ,
\end{equation}
where
\begin{align}
    \mathfrak{a} &= \frac{ 2 \mathfrak{f} }{ 3 } + \mathfrak{b} + \frac{ \mathfrak{c} ^2 }{ \mathfrak{b} }, &
    \mathfrak{b} &= \left[ - \frac{ \mathfrak{d} }{ 2 } + \sqrt{ \frac{ \mathfrak{d} ^2 }{ 4 } - \left( \mathfrak{c} ^2 \right) ^3 } \right] ^{ 1 / 3 }, &
    \mathfrak{c} ^2 &= \frac{ 4 \mathfrak{f} ^2 }{ 9 } - \frac{ \mathfrak{e} }{ 3 }, \\
    \mathfrak{d} &= - \frac{ 16 \mathfrak{f} ^3 }{ 27 } + \frac{ 2 \mathfrak{e} \mathfrak{f} }{ 3 } - 4 k_0 ^2 G_{\mathrm{P}} ^2 W ^2, &
    \mathfrak{e} &= W ^4 + 4 k_0 ^2 \epsilon W ^2 + \frac{ 4 \left( k_0 ^2 \epsilon - q_z ^2 \right) G_{\mathrm{P}} ^2 }{ \epsilon }, &
    \mathfrak{f} &= 2 \left( k_0 ^2 \epsilon - q_z ^2 \right) + W ^2 ,
\end{align}
and the branches of the square and cube roots are those giving $ \operatorname{Re} \left( \kappa_\pm \right) > 0 $.

For $ W \neq 0 $ the two inverse decay lengths are complex and are not related by complex conjugation. The hybridisation parameters are
\begin{equation}
    \Gamma_\pm = \frac{ G_{\mathrm{P}} \left( k_0 ^2 \epsilon - q_z ^2 \right) + \mathrm{i} W \kappa_\pm k_0 \epsilon }{ k_0 \epsilon \left( k_0 ^2 \epsilon - q_z ^2 + \kappa_\pm ^2 \right) } ,
\end{equation}
and the fields in the conductor read
\begin{align}
    \widetilde{\mathbf{E}}_\pm
    &=
    \frac{ \mathrm{i} \omega \mathcal{A}_\pm }{ q_z }
    \left\{
    \begin{bmatrix}
        - q_z \\[16pt]
        0 \\[16pt]
        0
    \end{bmatrix}
    +
    \mathrm{i} \Gamma_\pm
    \begin{bmatrix}
        0 \\[16pt]
        - \dfrac{ \left( \mathrm{i} \kappa_\pm + W / \Gamma_\pm \right) q_z ^2 }{ q_z ^2 - k_0 ^2 \epsilon } \\[16pt]
        q_z
    \end{bmatrix}
    \right\},
    &
    \widetilde{\mathbf{B}}_\pm
    & =
    \mathcal{A}_\pm
    \left\{
    \begin{bmatrix}
        0 \\[16pt]
        - \mathrm{i} q_z \\[16pt]
        \kappa_\pm
    \end{bmatrix}
    -
    \begin{bmatrix}
        \dfrac{ \mathrm{i} \Gamma_\pm \kappa_\pm k_0 ^2 \epsilon + W q_z ^2 }{ q_z ^2 - k_0 ^2 \epsilon } \\[16pt]
        0 \\[16pt]
        0
    \end{bmatrix}
    \right\} .
\end{align}

Matching the fields at $ y = 0 $ yields the secular equation
\begin{equation}
    \sum_{\sigma=\pm} \sigma \left[ \Gamma_{ - \sigma } \left( \kappa_{ \sigma } + \kappa_d \right) \left( \epsilon_d + \frac{ \epsilon \kappa_d \kappa_{ - \sigma } }{ q_z ^2 - k_0 ^2 \epsilon } \right) - \frac{ \mathrm{i} W \kappa_d q_z ^2 }{ k_0 ^2 \left( q_z ^2 - k_0 ^2 \epsilon \right) } \kappa_{ \sigma } \right] = 0 ,
\end{equation}
whose coefficients are complex for real $ \omega $ and $ q_z $, so that it does not satisfy the conjugation property discussed in Section~\ref{sec:PlanarPerp}. Its expansion at large wavevector gives the relation reported there, with an imaginary term of order $ 1 / | q_z | $.

\paragraph{AHE alone}

For $ W = 0 $ the imaginary term disappears, and the bulk dispersion relation becomes
\begin{equation}
    \left( q_z ^2 - \kappa_\pm ^2 - k_0 ^2 \epsilon \right) ^2 \epsilon - G_{\mathrm{P}} ^2 \left( k_0 ^2 \epsilon - q_z ^2 \right) = 0 ,
\end{equation}
which is again a quadratic equation for $ \kappa_\pm ^2 $, with roots
\begin{equation}
    \kappa_\pm = \sqrt{ q_z ^2 - k_0 ^2 \epsilon \pm \sqrt{ \frac{ G_{\mathrm{P}} ^2 }{ \epsilon } \left( k_0 ^2 \epsilon - q_z ^2 \right) } } ,
\end{equation}
and
\begin{equation}
    \Gamma_\pm = \pm \frac{ 1 }{ k_0 G_{\mathrm{P}} } \sqrt{ \frac{ G_{\mathrm{P}} ^2 }{ \epsilon } \left( k_0 ^2 \epsilon - q_z ^2 \right) }.
\end{equation}

Since $ \kappa_\pm $ depend on $ q_z $ only through $ q_z ^2 $ and $ \Gamma_\pm $ is odd in $ G_{\mathrm{P}} $, the dispersion relation is reciprocal and even in $ G_{\mathrm{P}} $, and its large-wavevector expansion coincides with that of the Faraday configuration.

\section{Technical Details on Cylindrical Interface}
\label{app:Cylindrical}

In the cylindrical geometry we use the coordinates $ \left( r , \phi , z \right) $ of Fig.~\ref{fig:figure1}(b), and consider fields proportional to $ \mathit{e} ^{ \mathrm{i} \left( m \phi + q_z z \right) } $. Since $ \mathbf{G} $ is along the axis, each harmonic $ m $ can be treated separately.

\subsection{Dielectric and conventional conductor eigenmodes}
\label{app:CylindricalDiel}

In an isotropic medium the longitudinal components $ E_z $ and $ B_z $ satisfy the modified Bessel equation,
\begin{equation}
  \left\{ \frac{ \mathrm{d} ^2 }{ \mathrm{d} r ^2 } + \frac{1}{r} \frac{ \mathrm{d} }{ \mathrm{d} r } - \left( \kappa_d ^2 + \frac{ m ^2 }{ r ^2 } \right) \right\}
  \begin{bmatrix}
    \widetilde{E}_{ z ,m } ^{ \mathrm{out} } \\
    \widetilde{B}_{ z ,m } ^{ \mathrm{out} }
  \end{bmatrix}
  = \mathbf{0} ,
\end{equation}
here written in the dielectric, with $ \kappa_d $ the same inverse decay length as in the planar geometry. The solutions that decay for $ r \to \infty $ are
\begin{equation}
  \widetilde{E}_{ z ,m } ^{ \mathrm{out} } = - \left( \frac{ \omega }{ c \, q_z } \right) \frac{ \kappa_d }{ q_z } \, \mathtt{E}_m ^{ \mathrm{out} } \, K_m \left( \kappa_d r \right) , \qquad \widetilde{B}_{ z ,m }^{ \mathrm{out} } = \frac{ \mathrm{i} \kappa_d }{ q_z } \, \mathtt{B}_m ^{ \mathrm{out} } \, K_m \left( \kappa_d r \right) ,
\end{equation}
where $ \mathtt{E}_m ^{ \mathrm{out} } $ and $ \mathtt{B}_m ^{ \mathrm{out} } $ are arbitrary amplitudes. The transverse components follow from Maxwell's equations,
\begin{align}
  E_{ r ,m } ^{ \mathrm{out} } &= \frac{ 1 }{ \mathrm{i} \kappa_d ^2 } \left[ q_z \frac{ \partial E_{ z ,m } ^{ \mathrm{out} } }{ \partial r } + \omega \frac{1}{r} \frac{ \partial B_{ z ,m } ^{ \mathrm{out} } }{ \partial \phi } \right] ,
  &
  E_{ \phi ,m } ^{ \mathrm{out} } &= \frac{ 1 }{ \mathrm{i} \kappa_d ^2 } \left[ \frac{ q_z }{ r } \frac{ \partial E_{ z ,m } ^{ \mathrm{out} } }{ \partial \phi } - \omega \frac{ \partial B_{ z ,m } ^{ \mathrm{out} } }{ \partial r } \right] ,
  \\[4pt]
  B_{ r ,m } ^{ \mathrm{out} } &= \frac{ 1 }{ \mathrm{i} \kappa_d ^2 } \left[ q_z \frac{ \partial B_{ z ,m } ^{ \mathrm{out} } }{ \partial r } - \frac{ \omega }{ c ^2 } \, \epsilon_d \frac{1}{r} \frac{ \partial E_{ z ,m } ^{ \mathrm{out} } }{ \partial \phi } \right] ,
  &
  B_{ \phi ,m } ^{ \mathrm{out} } &= \frac{ 1 }{ \mathrm{i} \kappa_d ^2 } \left[ \frac{ q_z }{ r } \frac{ \partial B_{ z ,m } ^{ \mathrm{out} } }{ \partial \phi } + \frac{ \omega }{ c ^2 } \, \epsilon_d \frac{ \partial E_{ z ,m } ^{ \mathrm{out} } }{ \partial r } \right] ,
\end{align}
so that the fields can be written compactly as
\begin{equation}
  \widetilde{\mathbf{E}}_m^{ \mathrm{out} } \left( r \right) = \frac{ \mathrm{i} \omega }{ c \, q_z } \left[ \mathtt{E}_m^{ \mathrm{out} } \, \mathbf{l}_m^{ \mathrm{out} } \left( r \right) + c \, \mathtt{B}_m^{ \mathrm{out} } \, \mathbf{t}_m^{ \mathrm{out} } \left( r \right) \right] , \qquad \widetilde{\mathbf{B}}_m^{ \mathrm{out} } \left( r \right) = \left( \frac{ \omega }{ c \, q_z } \right) ^2 \epsilon_d \frac{ \mathtt{E}_m^{ \mathrm{out} } }{ c } \, \mathbf{t}_m^{ \mathrm{out} } \left( r \right) + \mathtt{B}_m^{ \mathrm{out} } \, \mathbf{l}_m^{ \mathrm{out} } \left( r \right) ,
\end{equation}
where the vectors $ \mathbf{l}_m ^{\mathrm{out}} $ and $ \mathbf{t}_m ^{\mathrm{out}} $ read
\begin{equation}
  \mathbf{l}_m^{ \mathrm{out} } \left( r \right) =
  \begin{bmatrix}
    K_m' \left( \kappa_d r \right) \\[8pt]
    \dfrac{ \mathrm{i} m }{ \kappa_d r } K_m \left( \kappa_d r \right) \\[8pt]
    \dfrac{ \mathrm{i} \kappa_d }{ q_z } K_m \left( \kappa_d r \right)
  \end{bmatrix} ,
  \qquad
  \mathbf{t}_m^{ \mathrm{out} } \left( r \right) =
  \begin{bmatrix}
    \dfrac{ m }{ \kappa_d r } K_m \left( \kappa_d r \right) \\[8pt]
    \mathrm{i} \, K_m' \left( \kappa_d r \right) \\[8pt]
    0
  \end{bmatrix} .
\end{equation}

The amplitude vectors of Section~\ref{sec:Model} are therefore
\begin{align}
    \widetilde{\mathbf{E}}_m ^{\mathrm{I}} &= \frac{ \mathrm{i} \omega }{ c \, q_z } \, \mathtt{E}_m ^{\mathrm{out}} \, \mathbf{l}_m ^{\mathrm{out}} , &
    \widetilde{\mathbf{E}}_m ^{\mathrm{II}} &= \frac{ \mathrm{i} \omega }{ q_z } \, \mathtt{B}_m ^{\mathrm{out}} \, \mathbf{t}_m ^{\mathrm{out}} .
\end{align}

The expressions for the vectors inside the cylinder, $ \mathbf{l}_m^{ \mathrm{in} } \left( r \right) $ and $ \mathbf{t}_m^{ \mathrm{in} } \left( r \right) $, are obtained from the previous ones, $ \mathbf{l}_m^{ \mathrm{out} } \left( r \right) $ and $ \mathbf{t}_m^{ \mathrm{out} } \left( r \right) $, by the simple substitutions $ K_m \rightarrow I_m $, $ \epsilon_d \rightarrow \epsilon \left( \omega \right) $, and $ \kappa_d \rightarrow \kappa_\epsilon $.

Imposing the continuity of $ E_z $, $ E_\phi $, $ H_z $ and $ H_\phi $ at $ r = R $, we obtain the secular equation of a conventional cylindrical conductor,
\begin{equation}
  k_0 ^2 \left[ \frac{ \epsilon \left( \omega \right) }{ u } \frac{ I_m' \left( u \right) }{ I_m \left( u \right) } - \frac{ \epsilon_d }{ v } \frac{ K_m' \left( v \right) }{ K_m \left( v \right) } \right] \left[ \frac{1}{u} \frac{ I_m' \left( u \right) }{ I_m \left( u \right) } - \frac{1}{v} \frac{ K_m' \left( v \right) }{ K_m \left( v \right) } \right] - q_z ^2 \, m ^2 \left( \frac{1}{u ^2} - \frac{1}{v ^2} \right) ^2 = 0 ,
\end{equation}
where $ u = \kappa_\epsilon R $ and $ v = \kappa_d R $.

For $ m = 0 $ the last term vanishes, and the two factors give the TM and TE modes separately, while for $ m \neq 0 $ it couples the two polarisations.

\subsection{Anomalous conductor}
\label{app:CylindricalCond}

In the conductor we look for solutions in the form of Bessel functions of the first kind, with a transverse wavevector $ q_\perp $ that will be continued to imaginary values for the evanescent waves,
\begin{align}
    \widetilde{E}_{ r , m } ^{ \mathrm{in} } &\propto \alpha \, J_{m-1} \left( q_\perp r \right) + \beta \, J_{m+1} \left( q_\perp r \right) , &
    \widetilde{E}_{ \phi , m } ^{ \mathrm{in} } &\propto \mathrm{i} \left[ \alpha \, J_{m-1} \left( q_\perp r \right) - \beta \, J_{m+1} \left( q_\perp r \right) \right] , &
    \widetilde{E}_{ z , m } ^{ \mathrm{in} } &\propto \gamma \, J_{m} \left( q_\perp r \right).
\end{align}

The wave equation fixes
\begin{equation}
    \gamma = \frac{ \mathrm{i} q_\perp W }{ \left( k_0 ^2 \epsilon - q_\perp ^2 \right) } \left( \alpha + \beta \right) - \frac{ \mathrm{i} q_z q_\perp }{ \left( k_0 ^2 \epsilon - q_\perp ^2 \right) } \left( \alpha - \beta \right),
\end{equation}
and, using the recurrence relations~\eqref{eq:bessel-identities} of Appendix~\ref{app:MathBessel}, the transverse components give a linear system for $ \alpha \pm \beta $,
\begin{equation}
    \begin{bmatrix}
        \displaystyle \left( k_0 ^2 \epsilon - q_\perp ^2 - q_z ^2 \right) - \frac{ W G_z q_\perp ^2 }{ q_z k_0 \epsilon } &
        \displaystyle W q_z \left( 1 + \frac{ q_\perp ^2 }{ q_z ^2 } \right) - k_0 G_z \\[12pt]
        \displaystyle W q_z - k_0 G_z \left( 1 - \frac{ q_\perp ^2 }{ k_0 ^2 \epsilon } \right) &
        \displaystyle \left( k_0 ^2 \epsilon - q_\perp ^2 - q_z ^2 \right)
    \end{bmatrix}
    \begin{bmatrix}
        \alpha + \beta \\
        \alpha - \beta
    \end{bmatrix}
    = \mathbf{0} .
\end{equation}

Its determinant vanishes on the bulk dispersion relation with $ \mathbf{q} \cdot \mathbf{q} = q_z ^2 + q_\perp ^2 $ and $ \mathbf{q} \cdot \mathbf{G} = q_z G_z $,
\begin{equation}
    \left( q_z ^2 + q_{\perp,\pm} ^2 - k_0 ^2 \epsilon \right) ^2 \epsilon - G_z ^2 \left( k_0 ^2 \epsilon - q_{\perp,\pm} ^2 \right) - W ^2 \left( q_z ^2 + q_{\perp,\pm} ^2 \right) \epsilon + 2 k_0 \epsilon W G_z q_z = 0,
\end{equation}
which coincides with that of the planar Faraday configuration for $ q_{\perp,\pm} ^2 = - \kappa_\pm ^2 $. For each of the two roots, the ratio
\begin{equation}
    \Gamma_\pm = \left( \frac{ \alpha - \beta }{ \alpha + \beta } \right)_\pm ,
\end{equation}
plays the role of the hybridisation parameter of Appendix~\ref{app:PlanarCond}. Setting $ q_{\perp,\pm} = \mathrm{i} \kappa_\pm $ and using the analytic continuation $ J_m \left( \mathrm{i} \zeta \right) = \mathrm{i} ^m I_m \left( \zeta \right) $~\cite{NISTDLMF}, the fields of the two eigenmodes in the conductor are
\begin{align}
    \frac{ \widetilde{\mathbf{E}}_m ^\pm }{c}
    &=
    \frac{ \mathrm{i} \omega }{ c q_z }\mathcal{B}_m^\pm
    \left\{
    \begin{bmatrix}
        \displaystyle \frac{ m }{ \kappa_{\pm} r }I_{m} ^\pm \\[16pt]
        \displaystyle \mathrm{i} I_{m} ^{\pm \prime} \\[16pt]
        \displaystyle 0
    \end{bmatrix}
    +
    \Gamma_\pm
    \begin{bmatrix}
        \displaystyle I_{m} ^{\pm \prime} \\[16pt]
        \displaystyle \frac{ \mathrm{i} m }{ \kappa_{\pm} r } I_{m} ^\pm \\[16pt]
        \displaystyle \frac{\mathrm{i} \kappa_{\pm} }{ q_z } \Xi_\pm I_{m} ^\pm
    \end{bmatrix}
    \right\},
    &
    \widetilde{\mathbf{B}}_m ^\pm
    &=
    \mathcal{B}_m ^\pm
    \left\{
    \begin{bmatrix}
        \displaystyle I_{m} ^{\pm \prime} \\[16pt]
        \displaystyle \frac{ \mathrm{i} m }{ \kappa_{\pm} r } I_{m} ^\pm \\[16pt]
        \displaystyle \frac{\mathrm{i} \kappa_{\pm} }{ q_z } I_{m} ^\pm
    \end{bmatrix}
    +
    \Lambda_\pm
    \begin{bmatrix}
        \displaystyle \frac{ m }{ \kappa_{\pm} r } I_{m} ^\pm \\[16pt]
        \displaystyle \mathrm{i} I_{m} ^{\pm \prime} \\[16pt]
        \displaystyle 0
    \end{bmatrix}
    \right\},
\end{align}
where $ I_m ^\pm = I_m \left( \kappa_\pm r \right) $, the prime denotes the derivative with respect to the argument, and
\begin{align}
    \Xi_\pm &= \frac{ q_z ^2 - W q_z / \Gamma_\pm }{ k_0 ^2 \epsilon + \kappa_{\pm} ^2 }, &
    \Lambda_\pm &= \frac{ \Gamma_\pm k_0 ^2 \epsilon + \kappa_\pm ^2 W / q_z }{ k_0 ^2 \epsilon + \kappa_\pm ^2 }.
\end{align}

The fields have the same structure as in the dielectric, with the vectors $ \mathbf{l}_m ^{ \mathrm{in} } $ and $ \mathbf{t}_m ^{ \mathrm{in} } $ mixed by $ \Gamma_\pm $ and $ \Lambda_\pm $. Imposing the continuity of $ E_z $, $ E_\phi $, $ H_z $ and $ H_\phi $ at $ r = R $ and rescaling the amplitudes, we obtain the secular equation
\begin{equation}
    \begin{vmatrix}
        u_+ &
        u_- &
        v &
        0 \\[16pt]
        \displaystyle \Gamma_+ \Xi_+ u_+ &
        \displaystyle \Gamma_- \Xi_- u_- &
        0 &
        v \\[16pt]
        \displaystyle \Gamma_+ \frac{m}{u_+} +\frac{I_{m}^{+ \prime}}{I_{m}^+} &
        \displaystyle \Gamma_- \frac{m}{u_-} +\frac{I_{m}^{- \prime}}{I_{m}^-} &
        \displaystyle\frac{K_{m}'}{K_{m}} &
        \displaystyle\frac{m}{v} \\[16pt]
        \displaystyle \Lambda_+ \frac{I_{m}^{+ \prime}}{I_{m}^+} +\frac{m}{u_+} &
        \displaystyle \Lambda_- \frac{I_{m}^{- \prime}}{I_{m}^-} +\frac{m}{u_-} &
        \displaystyle\frac{m}{v} &
        \displaystyle\frac{k_0 ^2 \epsilon_d}{q_z ^2}\frac{K_{m}'}{K_{m}}
    \end{vmatrix}
    = 0,
\end{equation}
where $ u_\pm = \kappa_\pm R $, $ v = \kappa_d R $, and the Bessel functions are evaluated at $ r = R $.

For $ W = 0 $, $ \kappa_\pm $, $ \Gamma_\pm $, $ \Xi_\pm $ and $ \Lambda_\pm $ depend on $ q_z $ only through $ q_z ^2 $, and the propagation along the axis is reciprocal, while $ \Gamma_\pm $ and $ \Lambda_\pm $ are odd in $ G_z $, which couples $ m $ to $ G_z $. For $ \mathbf{G} = \mathbf{0} $, instead, $ \Gamma_\pm $ and $ \Lambda_\pm $ are odd in both $ q_z $ and $ W $, while $ \kappa_\pm $ and $ \Xi_\pm $ are even, and $ m $ is coupled to the linear momentum through the sign of $ q_z W $.

At large wavevector and fixed $ m $, the arguments $ u_\pm $ and $ v $ are large, and the expansions of the modified Bessel functions (see Appendix~\ref{app:MathBessel}), together with those of $ \kappa_\pm $, $ \kappa_d $ and $ \Gamma_\pm $, reduce the secular equation to
\begin{multline}
    \epsilon + \epsilon_d = \left[ \frac{ \epsilon + \epsilon_d }{ 2 R } - \frac{ \epsilon_d }{ R } + \frac{ G_z m }{ R k_0 } + \frac{ G_z W }{ 4 k_0 } \operatorname{sgn} \left( q_z \right) \right] \frac{ 1 }{ | q_z | } + \left[ \frac{ \epsilon_d W ^2 }{ 4 } - \epsilon_d ^2 k_0 ^2 - \frac{ G_z ^2 }{ 4 } - \frac{ G_z W }{ 8 R k_0 } \operatorname{sgn} \left( q_z \right) \right] \frac{ 1 }{ q_z ^2 } \\[8pt]
    + \left[ \frac{ G_z W \left( G_z ^2 + 4 \epsilon_d W ^2 - 32 \epsilon_d ^2 k_0 ^2 \right) }{ 64 \epsilon_d k_0 } \operatorname{sgn} \left( q_z \right) + \frac{ 2 G_z ^2 + \epsilon_d W ^2 - 8 \epsilon_d ^2 k_0 ^2 }{ 8 R } \right. \\[8pt]
    \left. + \frac{ W \left( 16 \epsilon_d k_0 m - 4 G_z m ^2 - 3 G_z \right) }{ 16 R ^2 k_0 } \operatorname{sgn} \left( q_z \right) + \frac{ \left( 4 m ^2 - 1 \right) \left( 3 \epsilon_d k_0 - G_z m \right) }{ 8 R ^3 k_0 } \right] \frac{ 1 }{ | q_z | ^3 } + \mathcal{O} \left( \frac{ \omega_p ^4 }{ c ^4 q_z ^4 } \right) .
\end{multline}

At large $ | m | $ and fixed $ q_z $, instead, the expansions of the Bessel functions at large order give
\begin{equation}
    \left( \epsilon + \epsilon_d \right) - \frac{ \operatorname{sgn} \left( m \right) \, G_z }{ k_0 } = - \frac{ R ^2 }{ 4 m ^2 k_0 } \left[ 2 \operatorname{sgn} \left( m \right) \, G_z q_z ^2 - G_z W q_z + \epsilon_d k_0 \left( 4 \epsilon_d k_0 ^2 - W ^2 \right) \right] + \mathcal{O} \left( \frac{ 1 }{ m ^3 } \right) .
\end{equation}
Inverted as in Appendix~\ref{app:MathAsym}, these two relations give the expansions of Section~\ref{sec:Cylindrical}.

\section{Mathematical Background}
\label{app:Math}

We collect here the identities used in the cylindrical geometry and the asymptotic expansion of the roots of the secular equations.

\subsection{Differential operators in cylindrical coordinates and Bessel functions}
\label{app:MathBessel}

For a vector field $ \mathbf{h} $ in cylindrical coordinates $ \left( r , \phi , z \right) $, with unit vectors $ \mathbf{e}_r $, $ \mathbf{e}_\phi $ and $ \mathbf{e}_z $, the differential operators used in Appendix~\ref{app:Cylindrical} read~\cite{Arfken2013}
\begin{align}
    \nabla \cdot \mathbf{h} &= \left( \frac{ 1 }{ r } + \frac{ \partial }{ \partial r } \right) h_r + \frac{ 1 }{ r } \frac{ \partial h_\phi }{ \partial \phi } + \frac{ \partial h_z }{ \partial z } , \\[6pt]
    \nabla \times \mathbf{h} &= \left( \frac{ 1 }{ r } \frac{ \partial h_z }{ \partial \phi } - \frac{ \partial h_\phi }{ \partial z } \right) \mathbf{e}_r + \left( \frac{ \partial h_r }{ \partial z } - \frac{ \partial h_z }{ \partial r } \right) \mathbf{e}_\phi + \frac{ 1 }{ r } \left[ \left( 1 + r \frac{ \partial }{ \partial r } \right) h_\phi - \frac{ \partial h_r }{ \partial \phi } \right] \mathbf{e}_z , \\[6pt]
    \nabla ^2 \mathbf{h} &= \left( \nabla ^2 h_r - \frac{ 1 }{ r ^2 } h_r - \frac{ 2 }{ r ^2 } \frac{ \partial h_\phi }{ \partial \phi } \right) \mathbf{e}_r + \left( \nabla ^2 h_\phi - \frac{ 1 }{ r ^2 } h_\phi + \frac{ 2 }{ r ^2 } \frac{ \partial h_r }{ \partial \phi } \right) \mathbf{e}_\phi + \left( \nabla ^2 h_z \right) \mathbf{e}_z , \\[6pt]
    \nabla ^2 h_{r,\phi,z} &= \left[ \left( \frac{ 1 }{ r } \frac{ \partial }{ \partial r } + \frac{ \partial ^2 }{ \partial r ^2 } \right) + \frac{ 1 }{ r ^2 } \frac{ \partial ^2 }{ \partial \phi ^2 } + \frac{ \partial ^2 }{ \partial z ^2 } \right] h_{r,\phi,z} .
\end{align}

For fields proportional to $ \mathit{e} ^{ \mathrm{i} \left( m \phi + q_z z \right) } $, the radial dependence of the solutions is given by Bessel functions~\cite{NISTDLMF}. Considering the Bessel functions of integer order $ \mathcal{V}_m = J_m $, for $ \mathfrak{s} = +1 $, and $ \mathcal{V}_m = I_m , \, \mathit{e} ^{ \mathrm{i} \pi m } K_m $, for $ \mathfrak{s} = -1 $, they satisfy the differential equation
\begin{equation}
  \zeta ^2 \mathcal{V}_m'' \left( \zeta \right) + \zeta \, \mathcal{V}_m' \left( \zeta \right) + \left( \mathfrak{s} \zeta ^2 - m ^2 \right) \mathcal{V}_m \left( \zeta \right) = 0 ,
\end{equation}
with the recurrence relations
\begin{equation}
  \mathcal{V}_{m-1} \left( \zeta \right) + \mathfrak{s} \, \mathcal{V}_{m+1} \left( \zeta \right) = \frac{ 2 m }{ \zeta } \, \mathcal{V}_m \left( \zeta \right) , \qquad
  \mathcal{V}_{m-1} \left( \zeta \right) - \mathfrak{s} \, \mathcal{V}_{m+1} \left( \zeta \right) = 2 \, \mathcal{V}_m' \left( \zeta \right) ,
  \label{eq:bessel-identities}
\end{equation}
which allow the components proportional to $ J_{m \mp 1} $ to be expressed through $ J_m $ and its derivative.

The logarithmic derivatives of $ I_m $ and $ K_m $ have the following asymptotic expansions. At large order and fixed argument they behave as
\begin{align}
    \frac{ I_m' \left( \zeta \right) }{ I_m \left( \zeta \right) } &= \frac{ | m | }{ \zeta } + \frac{ \zeta }{ 2 | m | } - \frac{ \zeta }{ 2 m ^2 } + \mathcal{O} \left( \frac{ 1 }{ | m | ^3 } \right) , &
    \frac{ K_m' \left( \zeta \right) }{ K_m \left( \zeta \right) } &= - \frac{ | m | }{ \zeta } - \frac{ \zeta }{ 2 | m | } - \frac{ \zeta }{ 2 m ^2 } + \mathcal{O} \left( \frac{ 1 }{ | m | ^3 } \right) ,
\end{align}
while at large argument and fixed order
\begin{align}
    \frac{ I_m' \left( \zeta \right) }{ I_m \left( \zeta \right) } &= 1 - \frac{ 1 }{ 2 \zeta } + \frac{ 4 m ^2 - 1 }{ 8 \zeta ^2 } + \frac{ 4 m ^2 - 1 }{ 8 \zeta ^3 } + \mathcal{O} \left( \frac{ 1 }{ \zeta ^4 } \right) , &
    \frac{ K_m' \left( \zeta \right) }{ K_m \left( \zeta \right) } &= - 1 - \frac{ 1 }{ 2 \zeta } - \frac{ 4 m ^2 - 1 }{ 8 \zeta ^2 } + \frac{ 4 m ^2 - 1 }{ 8 \zeta ^3 } + \mathcal{O} \left( \frac{ 1 }{ \zeta ^4 } \right) .
\end{align}

\subsection{Asymptotic expansion of the roots}
\label{app:MathAsym}

In the asymptotic regimes discussed above, the secular equation, divided by a nonvanishing overall factor, is equivalent, up to the order considered, to a relation of the form~\cite{BenderOrszag1999}
\begin{equation}\label{eq:asymptotic_form}
    \mathcal{F}_0 \left( \Omega \right) = \frac{ \mathcal{F}_1 \left( \Omega \right) }{ \xi } + \frac{ \mathcal{F}_2 \left( \Omega \right) }{ \xi ^2 } + \frac{ \mathcal{F}_3 \left( \Omega \right) }{ \xi ^3 } + \mathcal{O} \left( \frac{ 1 }{ \xi ^4 } \right),
\end{equation}
where $\Omega$ is the root to be found, here the frequency, $\xi \to \infty$ is the large parameter of the expansion (for instance $ q_z $ or $ m $), and the functions $\mathcal{F}_0, \dots, \mathcal{F}_3$ do not depend on $\xi$. At leading order the frequency is the root $\Omega_\infty$ of $\mathcal{F}_0 \left( \Omega_\infty \right) = 0$, assumed simple, $\mathcal{F}_0' \left( \Omega_\infty \right) \neq 0$. Since the right-hand side is of order $\xi ^{-1}$, so is the shift $\delta \Omega = \Omega - \Omega_\infty$. Expanding $\mathcal{F}_0$ to third order, $\mathcal{F}_1$ to second order and $\mathcal{F}_2$ to first order in $\delta \Omega$,
\begin{equation}
    \mathcal{F}_0' \, \delta \Omega + \frac{ 1 }{ 2 } \mathcal{F}_0'' \, \delta \Omega ^2 + \frac{ 1 }{ 6 } \mathcal{F}_0''' \, \delta \Omega ^3 = \frac{ 1 }{ \xi } \left( \mathcal{F}_1 + \mathcal{F}_1' \, \delta \Omega + \frac{ 1 }{ 2 } \mathcal{F}_1'' \, \delta \Omega ^2 \right)
    + \frac{ 1 }{ \xi ^2 } \left( \mathcal{F}_2 + \mathcal{F}_2' \, \delta \Omega \right) + \frac{ \mathcal{F}_3 }{ \xi ^3 } + \mathcal{O} \left( \frac{ 1 }{ \xi ^4 } \right),
\end{equation}
with all functions evaluated at $\Omega_\infty$. Inserting $\delta \Omega$ as a power series in $1 / \xi$ and matching the coefficients of $\xi ^{-1}$, $\xi ^{-2}$ and $\xi ^{-3}$ gives
\begin{equation}
    \Omega \left( \xi \right) = \Omega_\infty + \frac{ \mathcal{Y}_1 }{ \xi } + \frac{ \mathcal{Y}_2 }{ \xi ^2 } + \frac{ \mathcal{Y}_3 }{ \xi ^3 } + \mathcal{O} \left( \frac{ 1 }{ \xi ^4 } \right),
\end{equation}
with the coefficients
\begin{align}
    \mathcal{Y}_1 &= \frac{ \mathcal{F}_1 }{ \mathcal{F}_0' } , \\
    \mathcal{Y}_2 &= \frac{ 1 }{ \mathcal{F}_0' } \left( \mathcal{F}_2 + \mathcal{F}_1' \mathcal{Y}_1 - \frac{ 1 }{ 2 } \mathcal{F}_0'' \mathcal{Y}_1 ^2 \right) , \\
    \mathcal{Y}_3 &= \frac{ 1 }{ \mathcal{F}_0' } \left( \mathcal{F}_3 + \mathcal{F}_2' \mathcal{Y}_1 + \mathcal{F}_1' \mathcal{Y}_2 + \frac{ 1 }{ 2 } \mathcal{F}_1'' \mathcal{Y}_1 ^2 - \mathcal{F}_0'' \mathcal{Y}_1 \mathcal{Y}_2 - \frac{ 1 }{ 6 } \mathcal{F}_0''' \mathcal{Y}_1 ^3 \right) ,
\end{align}
each of which only requires the previous ones.

When $ \mathcal{F}_1 = 0 $, as for the CME alone and in the Voigt configuration, the expansion starts at order $ \xi ^{-2} $, and the coefficients $ \mathcal{Y}_2 $ and $ \mathcal{Y}_3 $ simplify accordingly.

\bibliographystyle{unsrturl}
\bibliography{references}

\begin{thebibliography}{10}

\bibitem{Barnes2003}
William Barnes, Alain Dereux, and Thomas Ebbesen.
\newblock Surface plasmon subwavelength optics.
\newblock {\em Nature}, 424(6950):824--830, 2003.
\newblock \href {https://doi.org/10.1038/nature01937} {\path{doi:10.1038/nature01937}}.

\bibitem{Maier2007}
Stefan~Alexander Maier.
\newblock {\em {Plasmonics: Fundamentals and Applications}}.
\newblock Springer, New York, 2007.

\bibitem{Yu2019}
Huakang Yu, Yusi Peng, Yong Yang, and Zhi-Yuan Li.
\newblock {Plasmon-Enhanced Light--Matter Interactions and Applications}.
\newblock {\em npj Computational Materials}, 5(1):45, 2019.
\newblock \href {https://doi.org/10.1038/s41524-019-0184-1} {\path{doi:10.1038/s41524-019-0184-1}}.

\bibitem{Gramotnev2010}
Dmitri~K. Gramotnev and Sergey~I. Bozhevolnyi.
\newblock {Plasmonics beyond the Diffraction Limit}.
\newblock {\em Nature Photonics}, 4(2):83--91, 2010.
\newblock \href {https://doi.org/10.1038/nphoton.2009.282} {\path{doi:10.1038/nphoton.2009.282}}.

\bibitem{Oulton2009}
Rupert~F. Oulton, Volker~J. Sorger, Thomas Zentgraf, Ren-Min Ma, Christopher Gladden, Lun Dai, Guy Bartal, and Xiang Zhang.
\newblock {Plasmon Lasers at Deep Subwavelength Scale}.
\newblock {\em Nature}, 461(7264):629--632, 2009.
\newblock \href {https://doi.org/10.1038/nature08364} {\path{doi:10.1038/nature08364}}.

\bibitem{Hill2009}
Martin~T. Hill et~al.
\newblock {Lasing in Metal-Insulator-Metal Sub-Wavelength Plasmonic Waveguides}.
\newblock {\em Optics Express}, 17(13):11107--11112, 2009.
\newblock \href {https://doi.org/10.1364/OE.17.011107} {\path{doi:10.1364/OE.17.011107}}.

\bibitem{Camley1987}
Robert~E. Camley.
\newblock {Nonreciprocal Surface Waves}.
\newblock {\em Surface Science Reports}, 7(3):103--187, 1987.
\newblock \href {https://doi.org/10.1016/0167-5729(87)90006-9} {\path{doi:10.1016/0167-5729(87)90006-9}}.

\bibitem{Caloz2018}
Christophe Caloz, Andrea Al{\`u}, Sergei Tretyakov, Dimitrios Sounas, Karim Achouri, and Zo{\'e}-Lise Deck-L{\'e}ger.
\newblock {Electromagnetic Nonreciprocity}.
\newblock {\em Physical Review Applied}, 10(4):047001, 2018.
\newblock \href {https://doi.org/10.1103/PhysRevApplied.10.047001} {\path{doi:10.1103/PhysRevApplied.10.047001}}.

\bibitem{ChiuQuinn1972}
K.~W. Chiu and J.~J. Quinn.
\newblock {Magnetoplasma Surface Waves in Metals}.
\newblock {\em Physical Review B}, 5(12):4707--4708, 1972.
\newblock \href {https://doi.org/10.1103/PhysRevB.5.4707} {\path{doi:10.1103/PhysRevB.5.4707}}.

\bibitem{Wallis1974}
R.~F. Wallis, J.~J. Brion, E.~Burstein, and A.~Hartstein.
\newblock {Theory of Surface Polaritons in Anisotropic Dielectric Media with Application to Surface Magnetoplasmons in Semiconductors}.
\newblock {\em Physical Review B}, 9(8):3424--3437, 1974.
\newblock \href {https://doi.org/10.1103/PhysRevB.9.3424} {\path{doi:10.1103/PhysRevB.9.3424}}.

\bibitem{Bliokh2018}
Konstantin~Y. Bliokh, Francisco~J. Rodr{\'i}guez-Fortu{\~n}o, Aleksandr~Y. Bekshaev, Yuri~S. Kivshar, and Franco Nori.
\newblock {Electric-Current-Induced Unidirectional Propagation of Surface Plasmon-Polaritons}.
\newblock {\em Optics Letters}, 43(5):963--966, 2018.
\newblock \href {https://doi.org/10.1364/OL.43.000963} {\path{doi:10.1364/OL.43.000963}}.

\bibitem{Kotov2018}
Oleg~V. Kotov and Yurii~E. Lozovik.
\newblock {Giant Tunable Nonreciprocity of Light in Weyl Semimetals}.
\newblock {\em Physical Review B}, 98(19):195446, 2018.
\newblock \href {https://doi.org/10.1103/PhysRevB.98.195446} {\path{doi:10.1103/PhysRevB.98.195446}}.

\bibitem{Guo2023}
Cheng Guo, Viktar~S. Asadchy, Bo~Zhao, and Shanhui Fan.
\newblock {Light Control with Weyl Semimetals}.
\newblock {\em eLight}, 3(1):2, 2023.
\newblock \href {https://doi.org/10.1186/s43593-022-00036-w} {\path{doi:10.1186/s43593-022-00036-w}}.

\bibitem{PecceiQuinn1977}
Roberto~D. Peccei and Helen~R. Quinn.
\newblock {CP Conservation in the Presence of Pseudoparticles}.
\newblock {\em Physical Review Letters}, 38(25):1440--1443, 1977.
\newblock \href {https://doi.org/10.1103/PhysRevLett.38.1440} {\path{doi:10.1103/PhysRevLett.38.1440}}.

\bibitem{Weinberg1978}
Steven Weinberg.
\newblock {A New Light Boson?}
\newblock {\em Physical Review Letters}, 40(4):223--226, 1978.
\newblock \href {https://doi.org/10.1103/PhysRevLett.40.223} {\path{doi:10.1103/PhysRevLett.40.223}}.

\bibitem{Wilczek1978}
Frank Wilczek.
\newblock {Problem of Strong P and T Invariance in the Presence of Instantons}.
\newblock {\em Physical Review Letters}, 40(5):279--282, 1978.
\newblock \href {https://doi.org/10.1103/PhysRevLett.40.279} {\path{doi:10.1103/PhysRevLett.40.279}}.

\bibitem{Sikivie1983}
Pierre Sikivie.
\newblock {Experimental Tests of the ``Invisible'' Axion}.
\newblock {\em Physical Review Letters}, 51(16):1415--1417, 1983.
\newblock \href {https://doi.org/10.1103/PhysRevLett.51.1415} {\path{doi:10.1103/PhysRevLett.51.1415}}.

\bibitem{Wilczek1987}
Frank Wilczek.
\newblock {Two Applications of Axion Electrodynamics}.
\newblock {\em Physical Review Letters}, 58(18):1799--1802, 1987.
\newblock \href {https://doi.org/10.1103/PhysRevLett.58.1799} {\path{doi:10.1103/PhysRevLett.58.1799}}.

\bibitem{Qi2008}
Xiao-Liang Qi, Taylor~L. Hughes, and Shou-Cheng Zhang.
\newblock {Topological Field Theory of Time-Reversal Invariant Insulators}.
\newblock {\em Physical Review B}, 78(19):195424, 2008.
\newblock \href {https://doi.org/10.1103/PhysRevB.78.195424} {\path{doi:10.1103/PhysRevB.78.195424}}.

\bibitem{Sekine2021}
Akihiko Sekine and Kentaro Nomura.
\newblock {Axion Electrodynamics in Topological Materials}.
\newblock {\em Journal of Applied Physics}, 129(14):141101, 2021.
\newblock \href {https://doi.org/10.1063/5.0038804} {\path{doi:10.1063/5.0038804}}.

\bibitem{Zyuzin2012}
A.~A. Zyuzin and A.~A. Burkov.
\newblock {Topological Response in Weyl Semimetals and the Chiral Anomaly}.
\newblock {\em Physical Review B}, 86(11):115133, 2012.
\newblock \href {https://doi.org/10.1103/PhysRevB.86.115133} {\path{doi:10.1103/PhysRevB.86.115133}}.

\bibitem{Chen2013}
Yige Chen, Si~Wu, and A.~A. Burkov.
\newblock {Axion Response in Weyl Semimetals}.
\newblock {\em Physical Review B}, 88(12):125105, 2013.
\newblock \href {https://doi.org/10.1103/PhysRevB.88.125105} {\path{doi:10.1103/PhysRevB.88.125105}}.

\bibitem{Fukushima2008}
Kenji Fukushima, Dmitri~E. Kharzeev, and Harmen~J. Warringa.
\newblock {Chiral Magnetic Effect}.
\newblock {\em Physical Review D}, 78(7):074033, 2008.
\newblock \href {https://doi.org/10.1103/PhysRevD.78.074033} {\path{doi:10.1103/PhysRevD.78.074033}}.

\bibitem{Kharzeev2014}
Dmitri~E. Kharzeev.
\newblock {The Chiral Magnetic Effect and Anomaly-Induced Transport}.
\newblock {\em Progress in Particle and Nuclear Physics}, 75:133--151, 2014.
\newblock \href {https://doi.org/10.1016/j.ppnp.2014.01.002} {\path{doi:10.1016/j.ppnp.2014.01.002}}.

\bibitem{Burkov2014}
A.~A. Burkov.
\newblock {Anomalous Hall Effect in Weyl Metals}.
\newblock {\em Physical Review Letters}, 113(18):187202, 2014.
\newblock \href {https://doi.org/10.1103/PhysRevLett.113.187202} {\path{doi:10.1103/PhysRevLett.113.187202}}.

\bibitem{Hofmann2016}
Johannes Hofmann and Sankar Das~Sarma.
\newblock {Surface Plasmon Polaritons in Topological Weyl Semimetals}.
\newblock {\em Physical Review B}, 93(24):241402, 2016.
\newblock \href {https://doi.org/10.1103/PhysRevB.93.241402} {\path{doi:10.1103/PhysRevB.93.241402}}.

\bibitem{Landau1984}
Lev~Davidovich Landau and Evgenii~Mikhailovich Lifshitz.
\newblock {\em {Electrodynamics of Continuous Media}}, volume~8 of {\em Course of Theoretical Physics}.
\newblock Pergamon Press, Oxford, 2 edition, 1984.

\bibitem{Ma2015}
Jing Ma and D.~A. Pesin.
\newblock {Chiral Magnetic Effect and Natural Optical Activity in Metals with or without Weyl Points}.
\newblock {\em Physical Review B}, 92(23):235205, 2015.
\newblock \href {https://doi.org/10.1103/PhysRevB.92.235205} {\path{doi:10.1103/PhysRevB.92.235205}}.

\bibitem{Vazifeh2013}
M.~M. Vazifeh and M.~Franz.
\newblock {Electromagnetic Response of Weyl Semimetals}.
\newblock {\em Physical Review Letters}, 111(2):027201, 2013.
\newblock \href {https://doi.org/10.1103/PhysRevLett.111.027201} {\path{doi:10.1103/PhysRevLett.111.027201}}.

\bibitem{Zhong2016}
Shudan Zhong, Joel~E. Moore, and Ivo Souza.
\newblock {Gyrotropic Magnetic Effect and the Magnetic Moment on the Fermi Surface}.
\newblock {\em Physical Review Letters}, 116(7):077201, 2016.
\newblock \href {https://doi.org/10.1103/PhysRevLett.116.077201} {\path{doi:10.1103/PhysRevLett.116.077201}}.

\bibitem{Son2013}
D.~T. Son and B.~Z. Spivak.
\newblock {Chiral Anomaly and Classical Negative Magnetoresistance of Weyl Metals}.
\newblock {\em Physical Review B}, 88(10):104412, 2013.
\newblock \href {https://doi.org/10.1103/PhysRevB.88.104412} {\path{doi:10.1103/PhysRevB.88.104412}}.

\bibitem{Li2016}
Qiang Li et~al.
\newblock {Chiral Magnetic Effect in ZrTe$_5$}.
\newblock {\em Nature Physics}, 12(6):550--554, 2016.
\newblock \href {https://doi.org/10.1038/nphys3648} {\path{doi:10.1038/nphys3648}}.

\bibitem{Hong2022}
Deog~Ki Hong, Sang~Hui Im, Kwang~Sik Jeong, and Dong-han Yeom.
\newblock {Detecting Axion Dark Matter with Chiral Magnetic Effects}.
\newblock {\em Physical Review D}, 110(5):055036, 2024.
\newblock \href {https://doi.org/10.1103/PhysRevD.110.055036} {\path{doi:10.1103/PhysRevD.110.055036}}.

\bibitem{Bose2026}
Debajit Bose, Prataya Chandra, Sudhansu~S. Mandal, and Tirtha~Sankar Ray.
\newblock {Probing Axion Dark Matter via the Chiral Magnetic Effect in Zero-Bias Weyl Semimetals}.
\newblock arXiv:2606.13595, 2026.
\newblock \href {https://doi.org/10.48550/arXiv.2606.13595} {\path{doi:10.48550/arXiv.2606.13595}}.

\bibitem{Irastorza2018}
Igor~G. Irastorza and Javier Redondo.
\newblock {New Experimental Approaches in the Search for Axion-Like Particles}.
\newblock {\em Progress in Particle and Nuclear Physics}, 102:89--159, 2018.
\newblock \href {https://doi.org/10.1016/j.ppnp.2018.05.003} {\path{doi:10.1016/j.ppnp.2018.05.003}}.

\bibitem{Marsh2019}
David J.~E. Marsh, Kin~Chung Fong, Erik~W. Lentz, Libor {\v{S}}mejkal, and Mazhar~N. Ali.
\newblock {Proposal to Detect Dark Matter Using Axionic Topological Antiferromagnets}.
\newblock {\em Physical Review Letters}, 123(12):121601, 2019.
\newblock \href {https://doi.org/10.1103/PhysRevLett.123.121601} {\path{doi:10.1103/PhysRevLett.123.121601}}.

\bibitem{SchutteEngel2021}
Jan Sch{\"u}tte-Engel et~al.
\newblock {Axion Quasiparticles for Axion Dark Matter Detection}.
\newblock {\em Journal of Cosmology and Astroparticle Physics}, 2021(08):066, 2021.
\newblock \href {https://doi.org/10.1088/1475-7516/2021/08/066} {\path{doi:10.1088/1475-7516/2021/08/066}}.

\bibitem{Tamaya2019}
Tomohiro Tamaya, Takeo Kato, Kota Tsuchikawa, Satoru Konabe, and Shiro Kawabata.
\newblock {Surface Plasmon Polaritons in Thin-Film Weyl Semimetals}.
\newblock {\em Journal of Physics: Condensed Matter}, 31(30):305001, 2019.
\newblock \href {https://doi.org/10.1088/1361-648X/ab17b3} {\path{doi:10.1088/1361-648X/ab17b3}}.

\bibitem{Bugaiko2020}
O.~V. Bugaiko, E.~V. Gorbar, and P.~O. Sukhachov.
\newblock {Surface Plasmon Polaritons in Strained Weyl Semimetals}.
\newblock {\em Physical Review B}, 102(8):085426, 2020.
\newblock \href {https://doi.org/10.1103/PhysRevB.102.085426} {\path{doi:10.1103/PhysRevB.102.085426}}.

\bibitem{OskouiAbdol2019}
S.~Oskoui~Abdol, A.~Soltani~Vala, and B.~Abdollahipour.
\newblock {Tunable Surface Plasmon Polaritons in a Weyl Semimetal Waveguide}.
\newblock {\em Journal of Physics: Condensed Matter}, 31(33):335002, 2019.
\newblock \href {https://doi.org/10.1088/1361-648X/ab217f} {\path{doi:10.1088/1361-648X/ab217f}}.

\bibitem{Peluso2025WG}
Marco Peluso, Alessandro De~Martino, Reinhold Egger, and Francesco Buccheri.
\newblock {Nonreciprocal Weyl Semimetal Waveguide}.
\newblock {\em Physical Review Research}, 7(2):023195, 2025.
\newblock \href {https://doi.org/10.1103/PhysRevResearch.7.023195} {\path{doi:10.1103/PhysRevResearch.7.023195}}.

\bibitem{Peluso2025NF}
Marco Peluso, Alessandro De~Martino, Reinhold Egger, and Francesco Buccheri.
\newblock {Angular-Momentum-Selective Nanofocusing with Weyl Semimetals}.
\newblock {\em Physical Review Research}, 7(4):043121, 2025.
\newblock \href {https://doi.org/10.1103/8c99-rvmm} {\path{doi:10.1103/8c99-rvmm}}.

\bibitem{OskouiAbdol2020}
S.~Oskoui~Abdol, B.~Abdollahipour, and A.~Soltani~Vala.
\newblock {Surface Plasmon Polaritons in a Waveguide Composed of Weyl Semimetals with Different Symmetries}.
\newblock {\em Journal of Physics D: Applied Physics}, 53(4):045105, 2020.
\newblock \href {https://doi.org/10.1088/1361-6463/ab52ee} {\path{doi:10.1088/1361-6463/ab52ee}}.

\bibitem{GomesMouraMelo2025}
Andr{\'e}~H. Gomes and Winder~A. Moura-Melo.
\newblock {Probing Emergent Axion Electrodynamics with Waveguide Apparatus}.
\newblock {\em Physical Review B}, 112(16):L161403, 2025.
\newblock \href {https://doi.org/10.1103/6v6c-4d7w} {\path{doi:10.1103/6v6c-4d7w}}.

\bibitem{AshcroftMermin}
Neil~W. Ashcroft and N.~David Mermin.
\newblock {\em {Solid State Physics}}.
\newblock Holt, Rinehart and Winston, New York, 1976.

\bibitem{Grosso2000}
G.~Grosso and G.~Pastori~Parravicini.
\newblock {\em {Solid State Physics}}.
\newblock Academic Press, San Diego, 2000.

\bibitem{Carroll1990}
Sean~M. Carroll, George~B. Field, and Roman Jackiw.
\newblock {Limits on a Lorentz- and Parity-Violating Modification of Electrodynamics}.
\newblock {\em Physical Review D}, 41(4):1231--1240, 1990.
\newblock \href {https://doi.org/10.1103/PhysRevD.41.1231} {\path{doi:10.1103/PhysRevD.41.1231}}.

\bibitem{Bender1998}
Carl~M. Bender and Stefan Boettcher.
\newblock {Real Spectra in Non-Hermitian Hamiltonians Having $\mathcal{PT}$ Symmetry}.
\newblock {\em Physical Review Letters}, 80(24):5243--5246, 1998.
\newblock \href {https://doi.org/10.1103/PhysRevLett.80.5243} {\path{doi:10.1103/PhysRevLett.80.5243}}.

\bibitem{ZyuzinZyuzin2015}
Alexander~A. Zyuzin and Vladimir~A. Zyuzin.
\newblock {Chiral Electromagnetic Waves in Weyl Semimetals}.
\newblock {\em Physical Review B}, 92(11):115310, 2015.
\newblock \href {https://doi.org/10.1103/PhysRevB.92.115310} {\path{doi:10.1103/PhysRevB.92.115310}}.

\bibitem{NISTDLMF}
{NIST Digital Library of Mathematical Functions}.
\newblock Release 1.2.8 of 2026-09-15.
\newblock F.~W.~J. Olver, A.~B. {Olde Daalhuis}, D.~W. Lozier, B.~I. Schneider, R.~F. Boisvert, C.~W. Clark, B.~R. Miller, B.~V. Saunders, H.~S. Cohl and M.~A. McClain, eds.
\newblock URL: \url{https://dlmf.nist.gov/}.

\bibitem{Arfken2013}
George~B. Arfken, Hans~J. Weber, and Frank~E. Harris.
\newblock {\em {Mathematical Methods for Physicists: A Comprehensive Guide}}.
\newblock Academic Press, Amsterdam, 7 edition, 2013.

\bibitem{BenderOrszag1999}
Carl~M. Bender and Steven~A. Orszag.
\newblock {\em {Advanced Mathematical Methods for Scientists and Engineers I: Asymptotic Methods and Perturbation Theory}}.
\newblock Springer, New York, 1999.

\end{thebibliography}

\end{document}